\documentclass[twocolumn]{aastex631}
\usepackage{amsmath}
\usepackage[T1]{fontenc}
\usepackage[utf8]{inputenc} 
\usepackage{upgreek}
\usepackage{breqn}

\newcommand{\PSUAA}{Department of Astronomy \& Astrophysics, 525 Davey Laboratory, The Pennsylvania State University, University Park, PA, 16802, USA}
\newcommand{\PSUCEHW}{Center for Exoplanets and Habitable Worlds, 525 Davey Laboratory, The Pennsylvania State University, University Park, PA, 16802, USA}
\newcommand{\PSETI}{Penn State Extraterrestrial Intelligence Center, 525 Davey Laboratory, The Pennsylvania State University, University Park, PA, 16802, USA}
\newcommand{\PSUICDS}{Institute for Computational and Data Sciences, The Pennsylvania State University, 525 Davey Laboratory, University Park, PA, 16802, USA}

\newcommand{\Penn}{Department of Physics and Astronomy, University of Pennsylvania, 209 S 33rd St, Philadelphia, PA 19104, USA}

\newcommand{\STScI}{Space Telescope Science Institute, 3700 San Martin Dr, Baltimore, MD 21218, USA}
\newcommand{\JHU}{Department of Physics and Astronomy, Johns Hopkins University, 3400 N Charles St, Baltimore, MD 21218, USA}
\newcommand{\GoddardESAL}{Exoplanets and Stellar Astrophysics Laboratory, NASA Goddard Space Flight Center, Greenbelt, MD 20771, USA}

\newcommand{\Macquarie}{Department of Physics and Astronomy, Macquarie University, Balaclava Road, North Ryde, NSW 2109, Australia }

\newcommand{\JPL}{Jet Propulsion Laboratory, California Institute of Technology, 4800 Oak Grove Drive, Pasadena, California 91109}

\newcommand{\UCI}{Department of Physics \& Astronomy, The University of California, Irvine, Irvine, CA 92697, USA}
\newcommand{\Carleton}{Carleton College, One North College St., Northfield, MN 55057, USA}
\newcommand{\Carnegie}{Earth and Planets Laboratory, Carnegie Institution for Science, 5241 Broad Branch Road, NW, Washington, DC 20015, USA}

\newcommand{\PSUCASt}{Center for Astrostatistics, 525 Davey Laboratory, The Pennsylvania State University, University Park, PA, 16802, USA}

\newcommand{\Princeton}{Department of Astrophysical Sciences, Princeton University, 4 Ivy Lane, Princeton, NJ 08540, USA}

\newcommand{\Wesleyan}{Astronomy Department and Van Vleck Observatory, Wesleyan University, 96 Foss Hill Drive, Middletown, CT 06459, USA}
\newcommand{\ETH}{ETH Zurich, Institute for Particle Physics \& Astrophysics, Zurich, Switzerland}

\newcommand{\Vanderbilt}{Department of Physics and Astronomy, Vanderbilt University, Nashville, TN 37235, USA}
\newcommand{\FlatironCCA}{Center for Computational Astrophysics, Flatiron Institute, 162 Fifth Avenue, New York, NY 10010, USA}
\newcommand{\Grenoble}{Univ. Grenoble Alpes, CNRS, IPAG, 414 rue de la Piscine, 38400, St-Martin d'Hères, France}
\newcommand{\UCSC}{Department of Astronomy and Astrophysics, University of California, Santa Cruz, CA 95064, USA}
\newcommand{\SETI}{SETI Institute, Carl Sagan Center, 339 Bernardo Ave, Suite 200, Mountain View, CA 94043, USA}
\newcommand{\Lehigh}{Department of Physics, Lehigh University, 16 Memorial Drive East, Bethlehem, PA, 18015, USA}

\newcommand{\teff}{T_{\rm eff}}

\newcommand{\rrev}{}
\newcommand{\rstrike}{\vphantom}

\received{2023 February 6}
\revised{2023 March 25}

\submitjournal{AJ}

\begin{document}

\title{A High-Eccentricity Warm Jupiter Orbiting TOI-4127}

\correspondingauthor{Arvind F.\ Gupta}
\email{arvind@psu.edu}

\author[0000-0002-5463-9980]{Arvind F.\ Gupta}
\affil{\PSUAA}
\affil{\PSUCEHW}

\author[0000-0002-0323-4828]{Jonathan M.\ Jackson}
\affil{\Wesleyan}
\affil{\PSUCEHW}

\author{Guillaume H\'ebrard}
\affil{Institut d'astrophysique de Paris, UMR7095 CNRS, Universit\'e Pierre \& Marie Curie,
98bis boulevard Arago, 75014 Paris, France}
\affil{Observatoire de Haute-Provence, CNRS, Universit\'e d'Aix-Marseille, 04870 Saint-Michel-l'Observatoire, France}

\author[0000-0002-9082-6337]{Andrea S.J.\ Lin}
\affil{\PSUAA}
\affil{\PSUCEHW}

\author[0000-0002-3481-9052]{Keivan G.\ Stassun}
\affil{\Vanderbilt}

\author[0000-0002-3610-6953]{Jiayin Dong} 
\altaffiliation{Flatiron Research Fellow} 
\affil{\FlatironCCA}

\author[0000-0001-6213-8804]{Steven Villanueva Jr.}
\altaffiliation{NASA Postdoctoral Program Fellow}
\affil{\GoddardESAL}

\author[0000-0003-2313-467X]{Diana Dragomir}
\affiliation{Department of Physics and Astronomy, The University of New Mexico, Albuquerque, NM 87106, USA}

\author[0000-0001-9596-7983]{Suvrath Mahadevan}
\altaffiliation{NEID Principal Investigator}
\affil{\PSUAA}
\affil{\PSUCEHW}
\affil{\ETH}

\author[0000-0001-6160-5888]{Jason T.\ Wright}
\altaffiliation{NEID Project Scientist}
\affil{\PSUAA}
\affil{\PSUCEHW}
\affil{\PSETI}


\author[0000-0003-3208-9815]{Jose M.\ Almenara}
\affil{\Grenoble}

\author[0000-0002-6096-1749]{Cullen H.\ Blake}
\affil{\Penn}

\author{Isabelle Boisse}
\affil{Aix Marseille Univ, CNRS, CNES, LAM, Marseille, France}

\author[0000-0002-6174-4666]{P\'ia Cort\'es-Zuleta}
\affil{Aix Marseille Univ, CNRS, CNES, LAM, Marseille, France}

\author[0000-0002-4297-5506]{Paul A.\ Dalba}
\altaffiliation{Heising-Simons 51 Pegasi b Postdoctoral Fellow}
\affiliation{\UCSC}
\affiliation{\SETI}

\author[0000-0001-9289-5160]{Rodrigo F.\ D\'iaz}
\affil{International Center for Advanced Studies (ICAS) and ICIFI (CONICET), ECyT-UNSAM, Campus Miguelete, 25 de Mayo y Francia, (1650) Buenos Aires, Argentina.}

\author[0000-0001-6545-639X]{Eric B.\ Ford}
\affil{\PSUAA}
\affil{\PSUCEHW}
\affil{\PSUICDS}
\affil{\PSUCASt}

\author[0000-0003-0536-4607]{Thierry Forveille}
\affil{\Grenoble}

\author[0000-0002-5665-1879]{Robert Gagliano}
\affil{Amateur Astronomer, Glendale, AZ 85308}

\author[0000-0003-1312-9391]{Samuel Halverson}
\affil{\JPL}

\author[0000-0002-2370-0187]{Neda Heidari}
\affil{Department of Physics, Shahid Beheshti University, Tehran, Iran.}
\affil{Laboratoire J.-L. Lagrange, Observatoire de la C\^ote d’Azur
(OCA), Universite de Nice-Sophia Antipolis (UNS), CNRS, Campus Valrose,
06108 Nice Cedex 2, France.}
\affil{Aix Marseille Univ, CNRS, CNES, LAM, Marseille, France.}

\author[0000-0001-8401-4300]{Shubham Kanodia}
\affil{\Carnegie}

\author[0000-0001-9129-4929]{Flavien Kiefer}
\affil{LESIA, Observatoire de Paris, Université PSL, CNRS, Sorbonne Université, Université Paris Cité, 5 place Jules Janssen, 92195 Meudon, France}

\author[0000-0001-9911-7388]{David w.\ Latham}
\affil{Center for Astrophysics ${\rm \mid}$ Harvard {\rm \&} Smithsonian, 60 Garden Street, Cambridge, MA 02138, USA}

\author[0000-0003-0241-8956]{Michael W.\ McElwain}
\affil{\GoddardESAL}

\author[0000-0002-4510-2268]{Ismael Mireles}
\affiliation{Department of Physics and Astronomy, The University of New Mexico, Albuquerque, NM 87106, USA}


\author[0000-0002-5463-9980]{Claire Moutou}
\affil{Universit\'e de Toulouse, CNRS, IRAP, 14 avenue Belin, 31400 Toulouse, France}

\author[0000-0002-3827-8417]{Joshua Pepper}
\affil{\Lehigh}

\author[0000-0003-2058-6662]{George R.\ Ricker}
\affil{Department of Physics and Kavli Institute for Astrophysics and Space Research, Massachusetts Institute of Technology, Cambridge, MA 02139, USA}

\author[0000-0003-0149-9678]{Paul Robertson}
\altaffiliation{NEID Project Scientist}
\affil{\UCI}

\author[0000-0001-8127-5775]{Arpita Roy}
\affil{\STScI}
\affil{\JHU}

\author[0000-0001-8355-2107]{Martin Schlecker}
\affiliation{Steward Observatory and Department of Astronomy, The University of Arizona, Tucson, AZ 85721, USA}

\author[0000-0002-4046-987X]{Christian Schwab}
\affil{\Macquarie}

\author[0000-0002-6892-6948]{S.\ Seager}
\affil{Department of Earth, Atmospheric, and Planetary Sciences, Massachusetts Institute of Technology, Cambridge, MA 02139, USA}
\affil{Department of Physics and Kavli Institute for Astrophysics and Space Research, Massachusetts Institute of Technology, Cambridge, MA 02139, USA}
\affil{Department of Aeronautics and Astronautics, Massachusetts Institute of Technology, Cambridge, MA 02139, USA}

\author[0000-0002-1836-3120]{Avi~Shporer}
\affiliation{Department of Physics and Kavli Institute for Astrophysics and Space Research, Massachusetts Institute of Technology, Cambridge, MA 02139, USA}

\author[0000-0001-7409-5688]{Guðmundur Stefánsson} 
\altaffiliation{NASA Sagan Fellow}
\affil{\Princeton}

\author[0000-0002-4788-8858]{Ryan C.\ Terrien}
\affil{\Carleton}

\author[0000-0002-8219-9505]{Eric B.\ Ting}
\affil{NASA Ames Research Center, Moffett Field, CA 94035, USA}

\author[0000-0002-4265-047X]{Joshua N.\ Winn}
\affil{\Princeton}

\author[0000-0002-1176-3391]{Allison Youngblood}
\affil{\GoddardESAL}


\begin{abstract}

We report the discovery of TOI-4127 b, a transiting, Jupiter-sized exoplanet on a long-period ($P = 56.39879^{+0.00010}_{-0.00010}$ d), high-eccentricity orbit around a late F-type dwarf star. This warm Jupiter was first detected and identified as a promising candidate from a search for single-transit signals in TESS Sector 20 data, and later characterized as a planet following two subsequent transits (TESS Sectors 26 and 53) and follow-up ground-based RV observations with the NEID and SOPHIE spectrographs. We jointly fit the transit and RV data to constrain the physical ($R_p = 1.096^{+0.039}_{-0.032} R_J$,  $M_p = 2.30^{+0.11}_{-0.11} M_J$) and orbital parameters of the exoplanet. Given its high orbital eccentricity ($e=0.7471^{+0.0078}_{-0.0086}$), TOI-4127 b is a compelling candidate for studies of warm Jupiter populations and of hot Jupiter formation pathways. We show that the present periastron separation of TOI-4127 b is too large for high-eccentricity tidal migration to circularize its orbit, and that TOI-4127 b is unlikely to be a hot Jupiter progenitor unless it is undergoing angular momentum exchange with an undetected outer companion. Although we find no evidence for an external companion, the available observational data are insufficient to rule out the presence of a perturber that can excite eccentricity oscillations and facilitate tidal migration.

\end{abstract}


\section{Introduction} \label{sec:intro}

One \rstrike{of the most}plausible explanation\rstrike{s} for the origin and existence of hot Jupiter exoplanets, or Jovian exoplanets with orbital periods $P\lesssim10$ days, is high-eccentricity tidal migration. In this scenario, giant exoplanets form at large orbital separations, analogous to the Jovian planets in our own Solar System, and then migrate inward to their present close-in, circular orbits following eccentricity excitation and subsequent tidal dissipation \citep{Wu2003}.
\rrev{This process can not be responsible for the origins of all hot Jupiter exoplanets, as it would preclude the population of systems with nearby companions \citep[e.g.,][]{Becker2015,Canas2019,Huang2020,Hord2022,Sha2022,Wu2023}. But it is likely to be one of several formation mechanisms at play \citep{Nelson2017,Dawson2018}.}
\rrev{The high-eccentricity tidal migration pathway makes several accurate predictions about the observed hot Jupiter population, including the}
\rstrike{This formation pathway accurately predicts the observed} pile-up of hot Jupiters with orbital periods of 3--4 days \citep{Wright2009,Santerne2016,Nelson2017}\rrev{, with the caveat that this may be partially attributable to selection effects \citep{Gaudi2005}}, as well as the \rrev{relative} dearth of systems with nearby companions \citep{Steffen2012}.

However, the transition from distant Jupiter analogs to close-in hot Jupiters \rrev{via high-eccentricity tidal migration} is not instantaneous. 
\rrev{Although multiple studies have demonstrated that this process can be accelerated in certain scenarios, leading to rapid migration for eccentric Jupiters with close periastron distances \citep{Wu2018,Rozner2022}, others show that}
\rstrike{If high-eccentricity tidal migration is the mechanism by}\rstrike{which a substantial fraction of hot Jupiters form,}we should \rstrike{also}expect to observe an intermediate, progenitor population of moderately- to highly-eccentric warm Jupiters, with orbital periods between 10 and 200 days \citep{Socrates2012,Dawson2015,Jackson2021,Jackson2022}.

The observed warm Jupiter eccentricity distribution extends out to the tidal limit \citep[e.g.,][]{Huang2016,Dong2021wj},
suggesting that the present orbits of some of these exoplanets have been shaped by the same types of excitations that lead to high-eccentricity tidal migration.
However, only two warm Jupiters, HD 80606 b \citep{Naef2001} and TOI-3362 b \citep{Dong2021}, have orbits that are sufficiently eccentric to ensure future tidal circularization and an ultimate fate as hot Jupiters.
For the remainder, the present, observed orbital architectures result in pericenter distances that are too great for tidal dissipation to be effective.
But these orbital architectures need not be static.
Several studies have shown that the orbital eccentricity of a planet can be excited to higher values via interactions with an external companion \citep[e.g., a wide stellar binary;][]{Mazeh1997,Holman1997,Innanen1997}.
This idea has been explored thoroughly in the context of hot and warm Jupiter formation. \citet{Socrates2012} and \citet{Petrovich2016} find that warm Jupiters may spend a fraction of their lives on a high-eccentricity tidal migration track if they are subject to secular eccentricity oscillations, through which they exchange angular momentum with a distant perturber to periodically reach higher eccentricities and tighter pericenter distances than presently observed.
Whether these eccentricity oscillations will be induced, and whether a given warm Jupiter can thus be classified as a hot Jupiter progenitor, hinges on the presence of a perturber. And perturbers have yet to be identified or ruled out in the majority of warm Jupiter systems.
In addition, for systems that have undergone high-eccentricity tidal migration, the dynamics of the migration pathway would preclude the existence of low-eccentricity warm Jupiters with small, nearby companions on co-planar orbits.
Yet many such systems have been discovered \citep[e.g.,][]{Tran2022}, suggesting that a significant number of warm Jupiters \rstrike{are}comprise a distinct population with a separate formation pathway.  
For warm Jupiters for which neither small, nearby companions nor distant, massive perturbers have been detected, it is as-yet undetermined which population they belong to, clouding our understanding of the frequency and demographics of hot Jupiter progenitors.
As such, further studies of eccentric warm Jupiters remain valuable, as these will allow us to better discern how some of these exoplanets reached their observed orbital states and where they might be headed in the future.

In this work, we present the discovery of a high-eccentricity warm Jupiter transiting the star TOI-4127. In Section \ref{sec:obs}, we first describe the photometric and spectroscopic data sets with which the exoplanet signal is detected and confirmed. We characterize the physical and orbital parameters of TOI-4127 b through joint analysis of TESS, NEID, and SOPHIE data in Section \ref{sec:analysis}. And we discuss the implications of these results for the dynamical origins of the system as well as prospects for future characterization efforts in Section \ref{sec:discussion}.

\section{Observations and Data Collection}\label{sec:obs}

\subsection{TESS Photometry}

\begin{figure*} 
    \centering
    \includegraphics[width=1.0\linewidth]{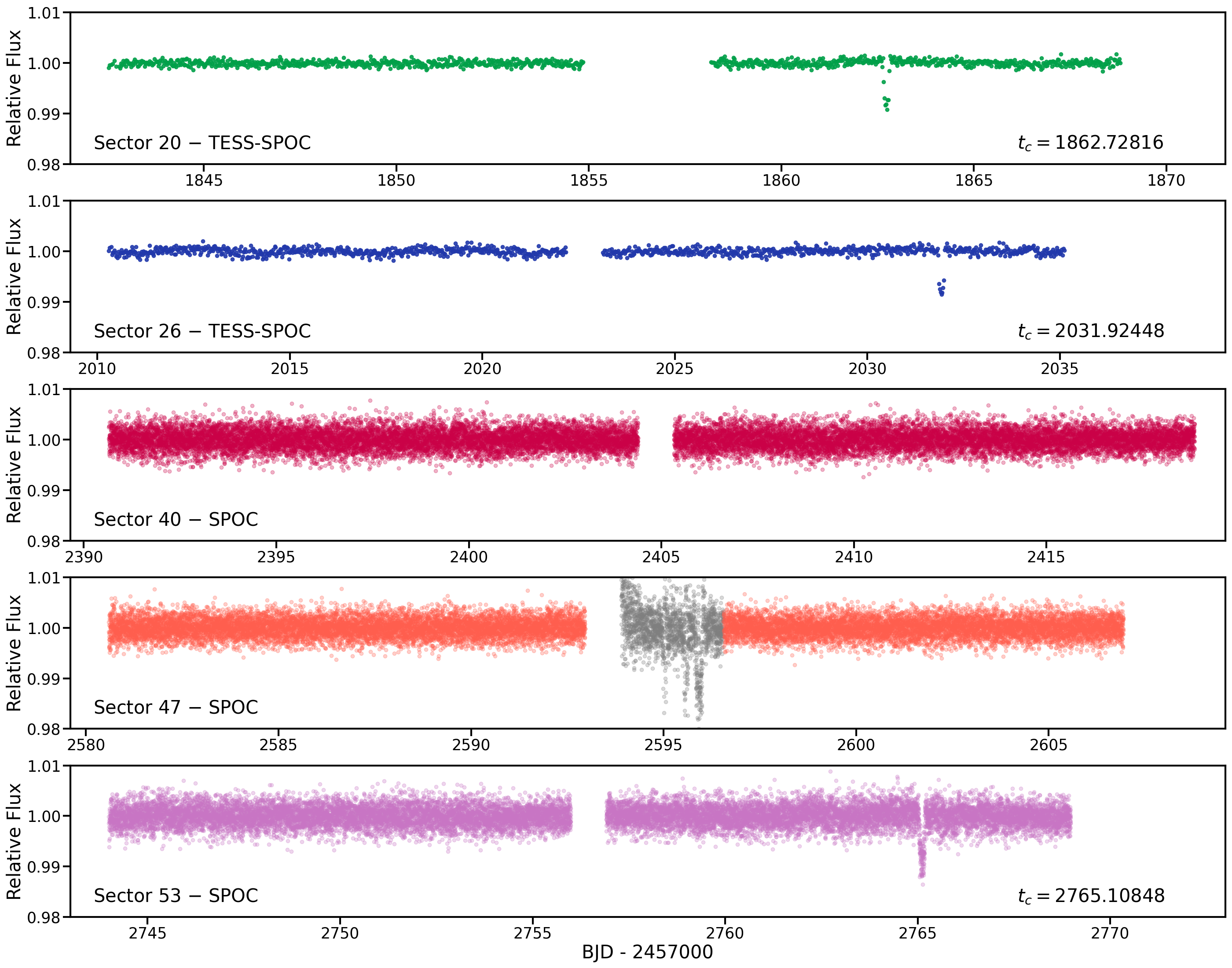}
    \caption{TESS PDCSAP photometry for TOI-4127. From top to bottom, we show long-cadence (30 minute) TESS-SPOC data for Sectors 20 and 26, and short-cadence (2 minute) SPOC data for Sectors 40, 47, and 53. We report the transit time in the lower right corner of each panel for the sectors in which high-quality transits were detected (Sectors 20, 26, and 53), and we show the SAP data for the low-quality Sector 47 transit in grey.}
    \label{fig:tess_lc}
\end{figure*}

TESS \citep{Ricker2015} observed TOI-4127 (TIC 141488193, Gaia DR3 1111639314047794944, 2MASS J07013709+7224538) in 5 non-contiguous sectors from 2019 December 19 (Sector 20) through 2022 July 9 (Sector 53). TOI 4127 was observed at a 30-minute cadence during the TESS primary mission in Sectors 20 and 26 and at a 2-minute cadence during the extended mission in Sectors 40, 47, and 53.  We show the pre-search data conditioned simple aperture photometry \citep[PDCSAP;][]{Smith2012,Stumpe2012,Stumpe2014} light curves from the Science Processing Operations Center \citep[SPOC;][]{Jenkins2016} for the extended mission TESS sectors and the TESS-SPOC \citep{Caldwell2020} light curves for the primary mission sectors in \autoref{fig:tess_lc}. 

A single transit was detected in the Sector 20 Quick Look Pipeline \citep[QLP;][]{Huang2020a,Huang2020b} extraction light curve by the TESS single-transit planet candidate working group using a custom search algorithm in May 2020, and a second transit with matching depth was detected in Sector 26 QLP data shortly thereafter.
Pixel-level light curves were generated using \textsf{eleanor} \citep{Feinstein2019} and analyzed to confirm that the transits were indeed associated with TOI-4127 and not a background or nearby source. 
The separation between the two transit midpoints in Sectors 20 and 26 was $\Delta t \approx 169$ days. But due to the gap in coverage between the two sectors, several period aliases were initially plausible until TESS returned to this star in Sectors 40, 47, and 53. Only a 56.4 day period alias remained consistent with the photometry following a third transit detection in Sector 53 and a non-detection in Sector 40. 

Although no transit is present in the PDCSAP light curve for Sector 47, we note that a transit would be expected to occur during the TESS observing baseline for this sector. In examining the light curve metadata, we find that the data in the expected transit window were flagged as low quality due to scattered light and thus excluded from analysis with the PDCSAP module. Using the simple aperture photometry (SAP) light curve, we confirm that a transit-like feature is indeed observed at a time consistent with the expected 56.4 day period (\autoref{fig:tess_lc}). However, given the poor quality of the data and the presence of uncharacterized light curve systematics, we restrict our analysis to the high-quality transits detected in Sectors 20, 26, and 53.

\subsection{NESSI Speckle Imaging}\label{sec:nessi}

\begin{figure} 
    \centering
    \includegraphics[width=1.0\linewidth]{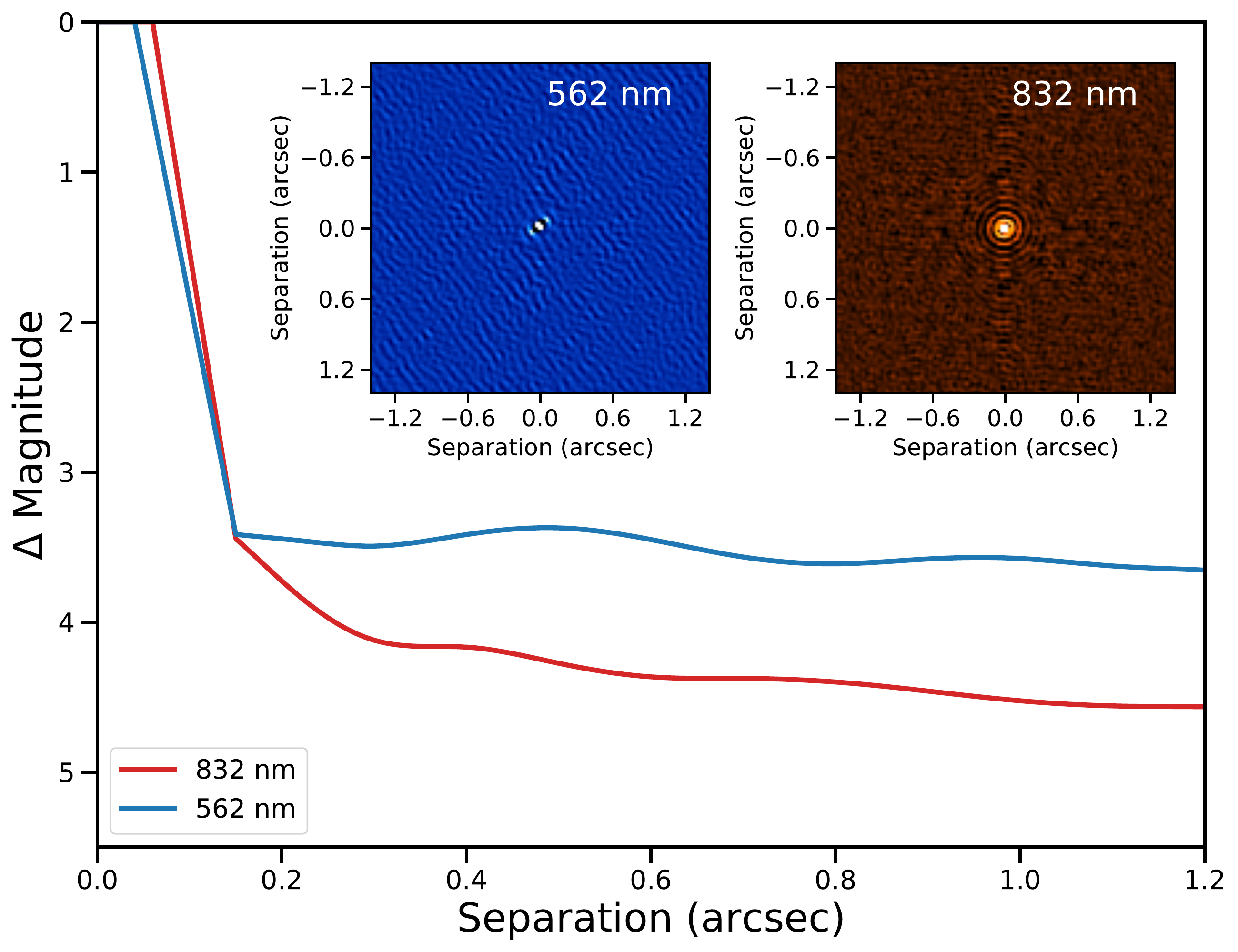}
    \caption{Reconstructed NESSI speckle images and 5$\sigma$ contrast curves for TOI-4127. Observations were taken simultaneously at 532 nm with the blue camera (left inset image) and at 832 nm with the red camera (right inset image). The contrast curves indicate the limiting magnitude difference at which bound or background stars could be detected for separations up to 1.2".}
    \label{fig:nessi}
\end{figure}

We observed TOI-4127 with the NN-Explore Exoplanet Stellar Speckle Imager \citep[NESSI;][]{Scott2018} on the WIYN\footnote{The WIYN Observatory is a joint facility of the NSF's National Optical-Infrared Astronomy Research Laboratory, Indiana University, the University of Wisconsin-Madison, Pennsylvania State University, the University of Missouri, the University of California-Irvine, and Purdue University.} 3.5m telescope at Kitt Peak National Observatory on the night of April 1, 2021.
Simultaneous, 1-minute sequences of 40 ms diffraction-limited exposures were taken in the 832 nm and 532 nm narrowband filters, and the speckle images were reconstructed using the procedures outlined in \citet{Howell2011}. We show the reconstructed images and accompanying $5\sigma$ contrast curves for both filters in \autoref{fig:nessi}. Our achieved contrast limits rule out the presence of faint stellar companions or background eclipsing binaries with $\Delta$mag $<3.3$ at 532 nm and $\Delta$mag$<3.9$ at 832 nm for separations $>0.2$".

\subsection{Spectroscopic follow-up observations}

Following the identification of transit signals in the first two sectors of TESS data for TOI-4127, we observed the star with the NEID \citep{Schwab2016} and SOPHIE \citep{Bouchy2006,Perruchot2008} spectrographs to measure the mass and orbit of the transiting exoplanet candidate. \rrev{Additional reconnaisance data from the Tillinghast Reflector Echelle Spectrograph \citep[TRES;][]{Furesz2008} were used to constrain the projected rotational velocity of the star.}

\subsection{TRES Spectroscopy}

\rrev{Four reconnaisance spectra were obtained with TRES, which is mounted on the 1.5m Tillinghast Reflector telescope at the Fred Lawrence Whipple Observatory (FLWO) atop Mount Hopkins, Arizona. TRES is an optical, fiber-fed echelle spectrograph with a wavelength range of 390 - 910 nm and a resolving power of R$\sim44,000$. The TRES spectra were first extracted and processed following the methods described in \citep{Buchhave2010} and the Stellar Parameter Classification \citep[SPC;][]{Buchhave2012,Buchhave2014} tool was then used to derive stellar parameters from the data. SPC cross correlates an observed stellar spectrum against a grid of spectra synthesized from Kurucz atmospheric models \citep{Kurucz1992} to derive the effective temperature, surface gravity, metallicity, and rotational velocity of the star.}

\subsubsection{NEID Spectroscopy}

NEID is a fiber-fed, red-optical (380 nm to 930 nm), environmentally stabilized \citep{Robertson2019} echelle spectrograph on the WIYN 3.5m telescope at Kitt Peak National Observatory. We obtained 12 NEID observations of TOI-4127 between February 6, 2021 and January 18, 2022 in the high resolution mode (resolving power of R $\sim 113,000$) with a median per-resolution-element S/N of $26.8$ at 550 nm. These data were processed with version $1.1.2$ of the NEID Data Reduction Pipeline\footnote{\url{https://neid.ipac.caltech.edu/docs/NEID-DRP/}} (DRP), which calculates RVs using the cross-correlation function \citep[CCF;][]{Baranne1996} method. 
Following the removal of one outlier point with S/N $= 6.22$, the CCF NEID RVs have a median single measurement precision of $8.09$ m~s$^{-1}$.
We also independently derive the RVs using a modified version of \textsf{SpEctrum Radial Velocity AnaLyser} \citep[\textsf{SERVAL};][]{Zechmeister2018}, a template-matching analysis method. Our modified \textsf{SERVAL} reduction follows the procedures described in \citet{Stefansson2022}, with the spectral range restricted to echelle orders 69-153 ($398$ nm $< \lambda < 895$ nm). In \autoref{tab:rvs_neid}, we list the NEID RVs from \textsf{SERVAL} (median single measurement precision $6.13$ m~s$^{-1}$), which performs slightly better than the DRP.

\begin{deluxetable}{rrr}
\tablecaption{NEID Radial Velocity Measurements for TOI-4127 \label{tab:rvs_neid}}
\tablehead{\colhead{Epoch (BJD)}&  \colhead{RV (m s$^{-1}$)}&
\colhead{$\sigma_{\rm RV}$ (m s$^{-1}$)}}
\startdata
2459251.69708 & 83.3 & 7.6 \\
2459256.70358 & 21.0 & 10.6 \\
2459273.67464 & -10.5 & 5.9 \\
2459481.94283 & 35.0 & 6.3 \\
2459489.94638 & -104.8 & 6.2 \\
2459538.99180 & -0.8 & 6.4 \\
2459539.72756 & -118.7 & 6.1 \\
2459568.02393 & 16.9 & 6.1 \\
2459586.98356 & 76.7 & 6.1 \\
2459596.66628 & -189.4 & 6.1 \\
2459597.62913 & -230.8 & 5.9 \\
\enddata
\end{deluxetable}

\subsubsection{SOPHIE spectroscopy}

Spectroscopic follow-up observations were also conducted with SOPHIE, a fiber-fed echelle spectrograph on the 1.93~m telescope at the Observatoire de Haute-Provence, France. A total of 21 high resolution (R $\sim75,000$) SOPHIE observations on the wavelength range $387-694$ nm were obtained between 2021 September 27 and 2022 December 7, with a median single measurement precision of $8$ m~s$^{-1}$. We removed two of the exposures that were noisy due to weather conditions (S/N per pixel at 550~nm below 20). The 19 remaining exposures have S/N between 20 and 36. As with NEID, the SOPHIE pipeline \citep{Bouchy2009} calculates RVs using the CCF method and using a G2-type numerical mask.
Seven of the SOPHIE exposures were corrected for spectral contamination from scattered moonlight, which has the potential to distort the CCF and shift the measured RV \citep{Roy2020}. Following the methods described in \citet{Hebrard2008} and \citet{Pollacco2008}, we calculate the CCF of both the stellar spectrum and a background sky spectrum, obtained simultaneously using a second fiber separated on sky by 2'. We then scale the CCFs by the relative throughputs of the two fibers and correct for scattered moonlight by subtracting the sky CCF from the stellar CCF.
We also calculate the CCF bisectors for each spectrum and find no correlation with the measured RVs, further supporting our argument that the transit signals in the TESS data were not due to blending with a background source.
The SOPHIE RVs are shown in \autoref{tab:rvs_sophie}.

\begin{deluxetable}{rrr}
\tablecaption{SOPHIE Radial Velocity Measurements for TOI-4127 \label{tab:rvs_sophie}}
\tablehead{\colhead{Epoch (BJD)}&  \colhead{RV (m s$^{-1}$)}&
\colhead{$\sigma_{\rm RV}$ (m s$^{-1}$)}}
\startdata
2459485.64778 & -213 & 7 \\
2459501.60591 & -49 & 10 \\
2459505.63623 & 18 & 9 \\
2459524.64269 & 48 & 9 \\
2459524.70562 & 69 & 8 \\
2459562.49480 & -5 & 6 \\
2459567.45250 & 22 & 9 \\
2459569.49023 & 44 & 8 \\
2459603.44302 & -86 & 8 \\
2459604.35637 & -60 & 7 \\
2459609.38166 & -33 & 11 \\
2459620.42060 & -6 & 9 \\
2459628.36524 & 23 & 10 \\
2459648.46673 & 66 & 7 \\
2459662.45629 & -52 & 7 \\
2459663.42345 & -69 & 8 \\
2459686.38970 & 74 & 9 \\
2459894.59742 & -6 & 7 \\
2459920.53907 & 78 & 7 \\
\enddata
\end{deluxetable}

\begin{figure*} 
    \centering
    \includegraphics[width=1.0\linewidth]{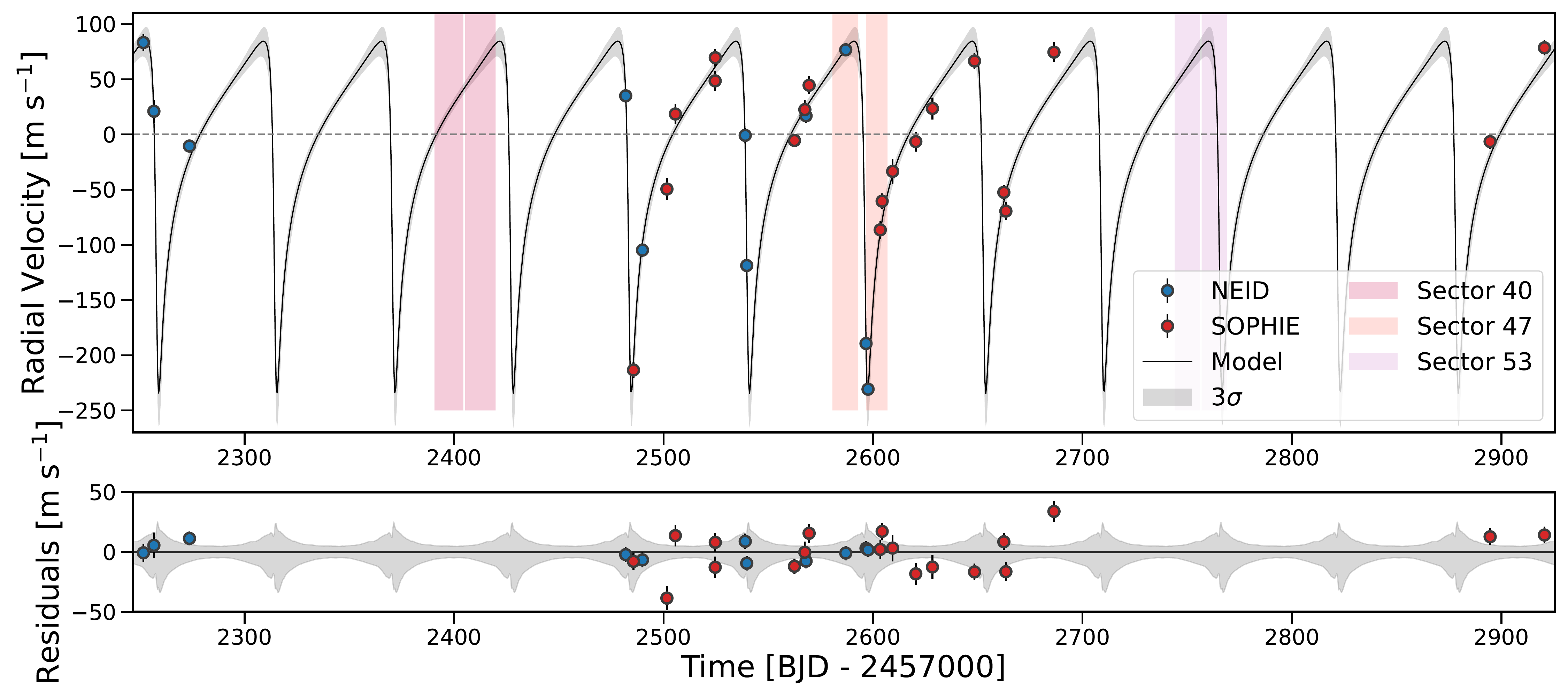}
    \caption{NEID and SOPHIE RV data for TOI-4127. We show the NEID data in blue and the SOPHIE data in red, as well as the median orbit fit, as determined in Section \ref{sec:analysis},  as a solid black line. We also highlight the times during which TESS was observing the target in Sectors 40 and 47, with colors matching the light curves in \autoref{fig:tess_lc}.}
    \label{fig:rvs}
\end{figure*}

As we show in \autoref{fig:rvs}, the NEID and SOPHIE RV measurements vary in phase with the TESS photometric ephemeris, and the semi-amplitude of the signal is consistent with that induced by a Jupiter-mass exoplanet. We conclude that the observed RV variations and transit signals for TOI-4127 are due to the presence of an eccentric giant exoplanet, TOI-4127 b.

\section{Analysis}\label{sec:analysis}

\subsection{Host Star Characterization}

We run a first order derivation of the stellar parameters using \textsf{SpecMatch-Emp} \citep{Yee2017}, comparing the observed NEID spectra to a library of High Resolution Echelle Spectrometer \citep[HIRES;][]{Vogt1994} reference spectra of well-characterized stars.
We first shift the observed NEID spectra to the barycentric rest frame and co-add these to produce a high-S/N master spectrum. We then run \textsf{SpecMatch-Emp} over the wavelength range 5052 -- 5807 \AA, broken up into $\sim$50 \AA\ chunks across the free spectral ranges of echelle orders 106-121.
\textsf{SpecMatch-Emp} calculates the best fit stellar parameters using a linear combination of the closest matching spectra in the library, with coefficients determined by a $\chi^2$ minimization of the composite library spectrum and target spectrum.
We achieve a good fit to the target spectrum across all orders. 
The derived stellar parameters, \rrev{$T_{\rm eff}$, $\log g$, and [Fe/H], are found to be consistent with the median values derived from the TRES spectra with SPC. The \textsf{SpecMatch-Emp} values} are listed in \autoref{tab:stellar_params} \rrev{along with the median TRES $v\sin i$}. 

\begin{figure} 
    \centering
    \includegraphics[width=0.74\linewidth,trim=70 80 90 90,clip,angle=90]{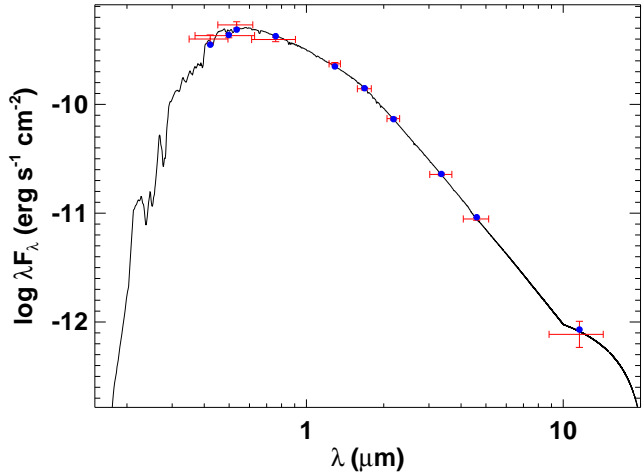}
    \caption{Spectral energy distribution of TOI-4127. Red symbols represent the observed photometric measurements, where the horizontal bars represent the effective width of the passband. Blue symbols are the model fluxes from the best-fit Kurucz atmosphere model (black).}
    \label{fig:sed}
\end{figure}

As an independent determination of the basic stellar parameters, we performed an analysis of the broadband spectral energy distribution (SED) of the star together with the {\it Gaia\/} EDR3 parallax \citep[with no systematic offset applied; see, e.g.,][]{StassunTorres:2021}, in order to determine an empirical measurement of the stellar radius, following the procedures described in \citet{Stassun:2016,Stassun:2017,Stassun:2018}. We pulled the $B_T V_T$ magnitudes from {\it Tycho-2} \citep[as reported by][]{Stassun2018}, the $JHK_S$ magnitudes from {\it 2MASS} \citep{Cutri2003}, the W1--W3 magnitudes from {\it WISE} \citep{Wright2010}, and the $G_{\rm BP} G_{\rm RP}$ magnitudes from {\it Gaia} \citep{GaiaCollaboration2022}. Together, the available photometry spans the full stellar SED over the wavelength range 0.4--10~$\mu$m (see \autoref{fig:sed}).

We performed a fit using Kurucz stellar atmosphere models \citep{Kurucz:1993}, with the effective temperature ($T_{\rm eff}$) and metallicity ([Fe/H]) adopted from the spectroscopically determined values. The remaining free parameter is the extinction $A_V$, which we limited to the maximum line-of-sight value from the Galactic dust maps of \citet{Schlegel:1998}. The resulting fit (Figure~\ref{fig:sed}) has a reduced $\chi^2$ of 1.1, with $A_V = 0.02 \pm 0.02$. Integrating the (de-reddened) model SED gives the bolometric flux at Earth, $F_{\rm bol} = 6.289 \pm 0.073 \times 10^{-10}$ erg~s$^{-1}$~cm$^{-2}$. Taking the $F_{\rm bol}$ with the {\it Gaia\/} parallax gives the bolometric luminosity directly as $L_{\rm bol} = 2.074 \pm 0.028$~L$_\odot$, and that together with $T_{\rm eff}$ gives the stellar radius, $R_\star = 1.293 \pm 0.050$~R$_\odot$. In addition, we can estimate the stellar mass from the empirical relations of \citet{Torres:2010}, giving $M_\star = 1.23 \pm 0.07$~M$_\odot$.

\begin{deluxetable*}{lrcc}
\tablecaption{Summary of Stellar Parameters for TOI-4127 \label{tab:stellar_params}}
\tablehead{\colhead{~~~Parameter}&  \colhead{Value}&
\colhead{Description}&
\colhead{Reference}}
\startdata
\multicolumn{4}{l}{\hspace{-0.2cm}Alternate Identifiers:}  \\
~~~TIC & 141488193 & TESS Input Catalog & Stassun \\
~~~Gaia DR3 &  1111639314047794944 & --- & Gaia DR3\\
~~~2MASS &  J07013709+7224538 & --- & 2MASS\\
\multicolumn{4}{l}{\hspace{-0.2cm} Coordinates and Parallax:} \\
~~~$\alpha_{\mathrm{J2016}}$ &    07:01:37.09 & Right Ascension (RA) & Gaia DR3\\
~~~$\delta_{\mathrm{J2016}}$ &   +72:24:53.75 & Declination (Dec) & Gaia DR3\\
~~~$\varpi$  & $3.07 \pm 0.02$ &  Parallax (mas) & Gaia DR3 \\
\multicolumn{4}{l}{\hspace{-0.2cm} Broadband photometry:}  \\
~~~$B_T$  &  $12.020 \pm 0.151$ & Tycho$-B_T$ & Stassun \\
~~~$V_T$  &  $11.441 \pm 0.010$ & Tycho$-V_T$ & Stassun \\
~~~TESS &  $11.0383 \pm 0.0061$ & --- & Stassun \\
~~~$G$  &  $11.4090 \pm 0.0003$ & Gaia & Gaia DR3\\
~~~$B_p$ & $11.6841 \pm 0.0003$ & Gaia & Gaia DR3\\
~~~$R_p$  &  $10.9814 \pm 0.0004$ & Gaia & Gaia DR3 \\
~~~$J$  &  $10.538 \pm 0.023$ & --- & 2MASS \\
~~~$H$  &  $10.308 \pm 0.023$ & --- & 2MASS \\
~~~$K_s$  &  $10.245 \pm 0.016$ & --- & 2MASS \\
~~~$W_1$  &  $10.206 \pm 0.022$ & --- & WISE \\
~~~$W_2$  &  $10.249 \pm 0.020$ & --- & WISE \\
~~~$W_3$  &  $10.277 \pm 0.051$ & --- & WISE \\
~~~$W_4$  &  $8.921$ & --- & WISE \\
\multicolumn{4}{l}{\hspace{-0.2cm} Derived Stellar Parameters:}\\
~~~$\teff$ & $6096\pm115$ & Effective Temperature (K) & This Work$^{a}$\\
~~~$\log g$ & $4.26\pm0.14$ & Surface Gravity ($\log$ (cm s$^{-2}$)) &  This Work$^{a}$\\
~~~[Fe/H] & $0.14\pm0.12$ & Metallicity (dex) &  This Work$^{a}$\\
~~~$v \sin i$ & $7.7\pm0.5$ & \rrev{Projected Rotational Velocity (km s$^{-1}$)} &  This Work$^{b}$\\
~~~$M_\star$ & $1.230\pm0.070$ & Stellar Mass (${\rm M}_\odot$)  &  This Work$^{c}$\\
~~~$R_\star$ & $1.293\pm0.050$ & Stellar Radius (${\rm R}_\odot$)   &  This Work$^{c}$\\
~~~$L_{\star,{\rm bol}}$ & $2.072\pm0.028$  &  Bolometric Luminosity (${\rm L}_\odot$) & This Work$^{c}$\\
~~~$F_{\star,{\rm bol}}$ & $6.289\pm0.073\times10^{-10}$  &  Bolometric Flux (erg~s$^{-1}$~cm$^{-2}$) & This Work$^{c}$\\
~~~Age & $4.8\pm2.1$ & Stellar Age (Gyr) &  This Work$^{c}$\\
~~~$A_V$ & $0.02\pm0.02$ & Extinction (mag) &  This Work$^{c}$\\
\enddata
\tablenotetext{a}{Parameters from \textsf{SpecMatch-Emp} analysis \rrev{of NEID spectra}}
\tablenotetext{b}{\rrev{Parameters from SPC analysis of TRES spectra}}
\tablenotetext{c}{Parameters from SED analysis}
\tablenotetext{}{References are: Stassun \citep{Stassun2018}, Gaia DR3 \citep{GaiaCollaboration2022}, 2MASS \citep{Cutri2003}, WISE \citep{Wright2010}}
\end{deluxetable*}

\subsection{Joint transit + RV modeling}\label{sec:jointfit}

We use the \textsf{exoplanet} package \citep{Foreman-Mackey2021} to jointly model the measured photometric and radial velocity signals for TOI-4127, with Gaussian priors on the stellar mass and radius imposed by the results of our \textsf{SpecMatch-Emp} and SED analysis, and flat priors on the exoplanet orbital period ($P$), transit depth ($\delta$), and transit impact parameter ($b$).
We model the orbit using a full Keplerian with uniform priors on exoplanet mass ($M_p$), orbital period ($P$), and transit impact parameter ($b$), and we sample the eccentricity and argument of periastron on the unit disk in $\sqrt{e}\sin\omega$-$\sqrt{e}\cos\omega$ space. We elect not to place informative priors on the eccentricity, as previous work has shown that there is no strongly favored eccentricity distribution for the known warm Jupiter population \citep{Dong2021wj}.
The orbital parameters are fit simultaneously to both the RV and photometric data sets.
For the photometry, we also place uniform priors on the transit depth ($\delta$) and we include a quadratic limb darkening law, re-parameterized as in \citet{Kipping2013}, and a Matern-3/2 Gaussian process kernel \citep{Foreman-Mackey2017,Foreman-Mackey2018} for the variability of the underlying light curve.
We fit separate jitter terms ($\sigma$) and offset terms ($\gamma$) for NEID and SOPHIE, and we allow for a long term, linear background trend.

\begin{figure} 
    \centering
    \includegraphics[width=1.0\linewidth]{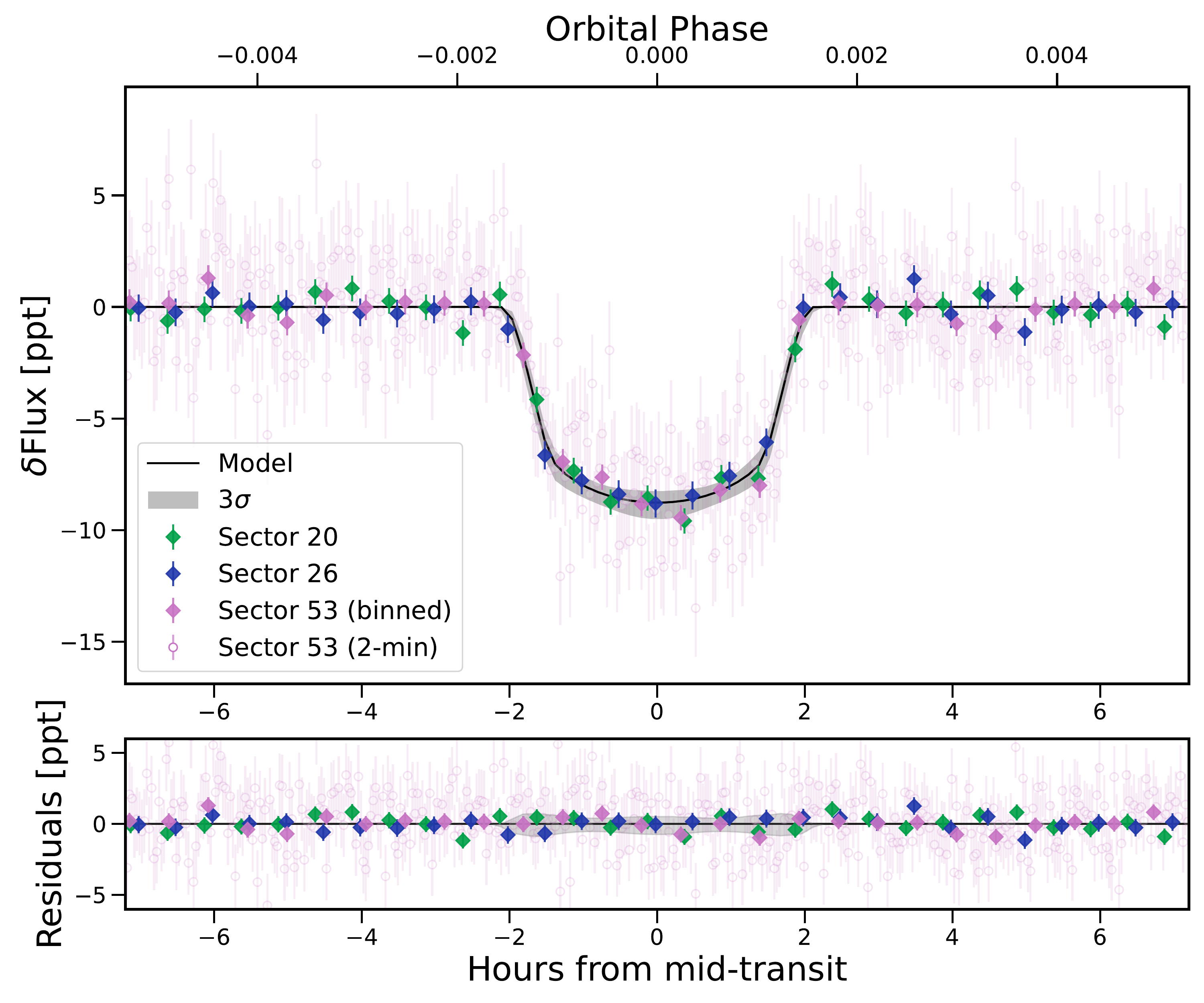}
    \caption{In-transit TESS data for TOI-4127 b, folded to the fitted orbital period. In the top panel, we plot the Sector 20 and Sector 26 data in green and blue, respectively, and the Sector 53 data in pink, to match the light curves shown in \autoref{fig:tess_lc}. We include both the binned (30 minute cadence) and unbinned (2 minute cadence) Sector 53 transit data. We show the best-fit model in black and $3\sigma$ uncertainies in grey, and we plot the residuals to this fit in the bottom panel.}
    \label{fig:transit_fit}
\end{figure}

\begin{figure} 
    \centering
    \includegraphics[width=1.0\linewidth]{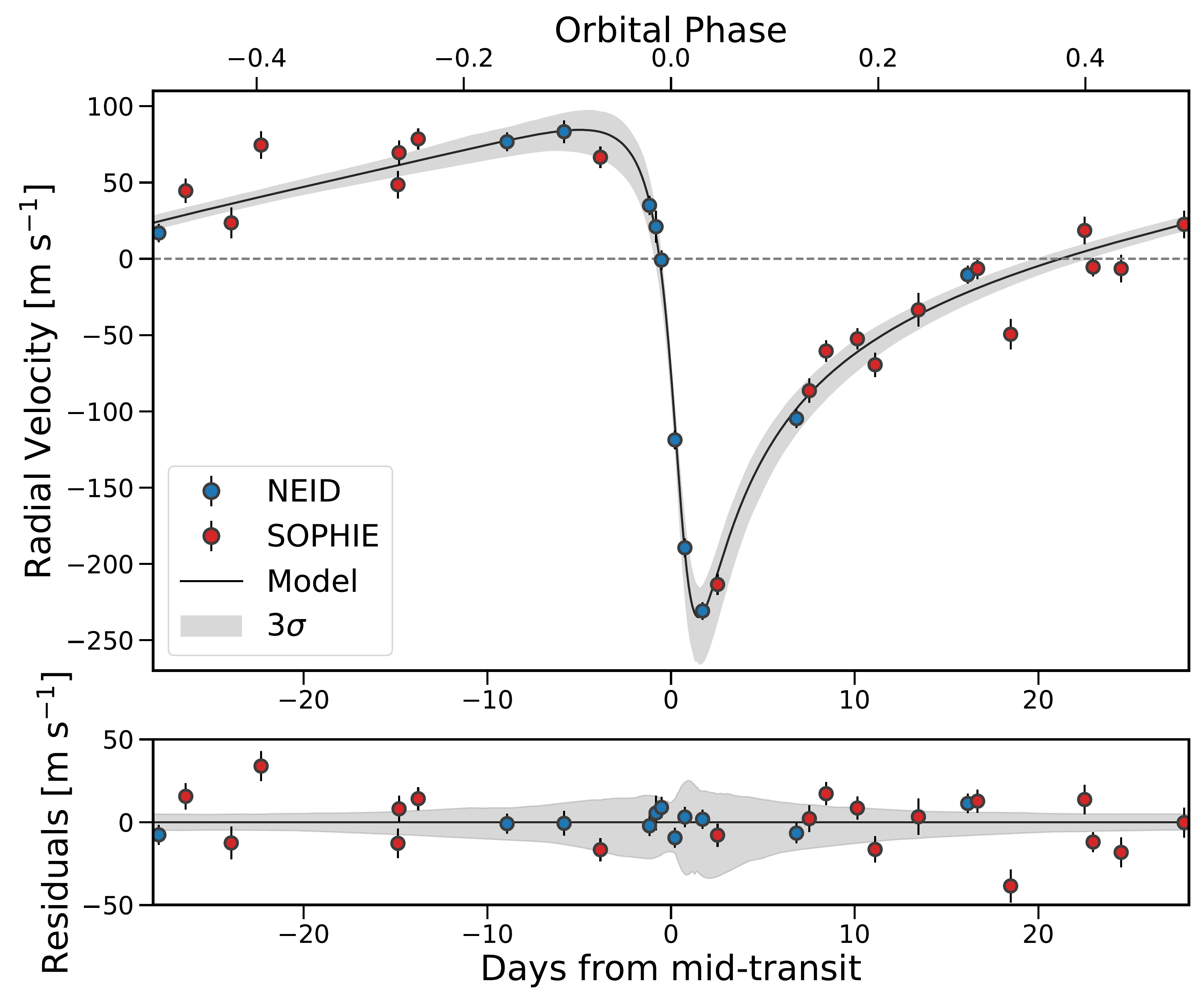}
    \caption{Phase-folded RV data for TOI-4127 b. We show the NEID data (blue) and SOPHIE data (red) as well as the best-fit model (black) and $3\sigma$ uncertainies (grey) in the top panel, and the residuals to the fit in the bottom panel.}
    \label{fig:rv_fit}
\end{figure}

From the best-fit solution, we confirm that TOI-4127 b is a warm Jupiter on an eccentric orbit. The fitted orbital period is $P = 56.39879^{+0.00010}_{-0.00010}$ days, which is in good agreement with the initial estimate from observed transit separation, and we report a relatively high eccentricity of $e = 0.7471^{+0.0078}_{-0.0086}$. The best-fit companion mass and radius are $M_p = 2.30^{+0.11}_{-0.11} M_J$ and $R_p = 1.096^{+0.039}_{-0.032} R_J$, consistent with a gas giant exoplanet.
The transit and phase-folded orbit are shown in Figures \ref{fig:transit_fit} and \ref{fig:rv_fit}, respectively, and we list the fitted parameters in \autoref{tab:derived_params}.

\begin{deluxetable*}{lrc}
\tablecaption{Derived parameters for TOI-4127 b \label{tab:derived_params}}
\tablehead{\colhead{~~~Parameter}&  \colhead{Value}&
\colhead{Description}}
\startdata
\multicolumn{3}{l}{\hspace{-0.2cm} Exoplanet Parameters:}  \\
$P$ & $56.39879^{+0.00010}_{-0.00010}$ & Orbital Period (days)\\
$T_C$ & $2458862.72816^{+0.00096}_{-0.00100}$ & Time of Conjunction (BJD)\\
$e$ & $0.7471^{+0.0078}_{-0.0086}$ & Eccentricity\\
$\omega$ & $129.6^{+2.0}_{-2.0}$ & Argument of Periastron ($^\circ$)\\
$i$ & $89.30^{+0.46}_{-0.60}$ & Inclination ($^\circ$)\\
$b$ & $0.17^{+0.14}_{-0.11}$ & Transit Impact Parameter\\
$K$ & $160.1^{+3.7}_{-3.8}$ & RV Semi-amplitude (m s$^{-1}$)\\
$a$ & $0.3081^{+0.0055}_{-0.0058}$ & Semi-major Axis (AU)\\
$u_1$ & $0.49^{+0.16}_{-0.18}$ & Linear Limb\\
$u_2$ & $-0.05^{+0.36}_{-0.19}$ & Quadratic Limb\\
$\delta$ & $7.46^{+0.25}_{-0.27}$ & Transit Depth (ppt)\\
$M_p$ & $2.30^{+0.11}_{-0.11}$ & Planet Mass ($M_J$)\\
$R_p$ & $1.096^{+0.039}_{-0.032}$ & Planet Radius ($R_J$)\\
$\rho_p$ & $2.17^{+0.22}_{-0.24}$ & Planet Density (g cm$^{-3}$)\\
$\rho_\star$ & $0.775^{+0.066}_{-0.070}$ & Stellar Density (g cm$^{-3}$)\\
$S$ & $22.3^{+1.5}_{-1.2}$ & Insolation ($S_\oplus$) \\
$T_{eq}$ & $605.1^{+9.6}_{-8.2}$ & Equilibrium Temperature (K)\\
\multicolumn{3}{l}{\hspace{-0.2cm} Instrument and Model Parameters:}  \\
$\sigma_{\rm phot, TESS}$ & $520^{+8}_{-7}$ & TESS Photometric Jitter (ppm)\\
$\sigma_{\rm GP}$ & $231^{+15}_{-17}$ & Covariance Amplitude (ppm)\\
$\rho_{\rm GP}$ & $0.298^{+0.091}_{-0.067}$ & Covariance Timescale (days)\\
$\sigma_{\rm RV, NEID}$ & $5.6^{+3.9}_{-3.2}$ & NEID RV Jitter (m~s$^{-1}$)\\
$\gamma_{\rm RV, NEID}$ & $-36.2916^{+0.0036}_{-0.0035}$ & NEID Systemic Velocity (km~s$^{-1}$)\\
$\sigma_{\rm RV, SOPHIE}$ & $15.1^{+2.8}_{-3.0}$ & SOPHIE RV Jitter (m~s$^{-1}$)\\
$\gamma_{\rm RV, SOPHIE}$ & $-36.814^{+0.004}_{-0.004}$ & SOPHIE Systemic Velocity (km~s$^{-1}$)\\
\enddata
\end{deluxetable*}

\begin{figure} 
    \centering
    \includegraphics[width=1.0\linewidth]{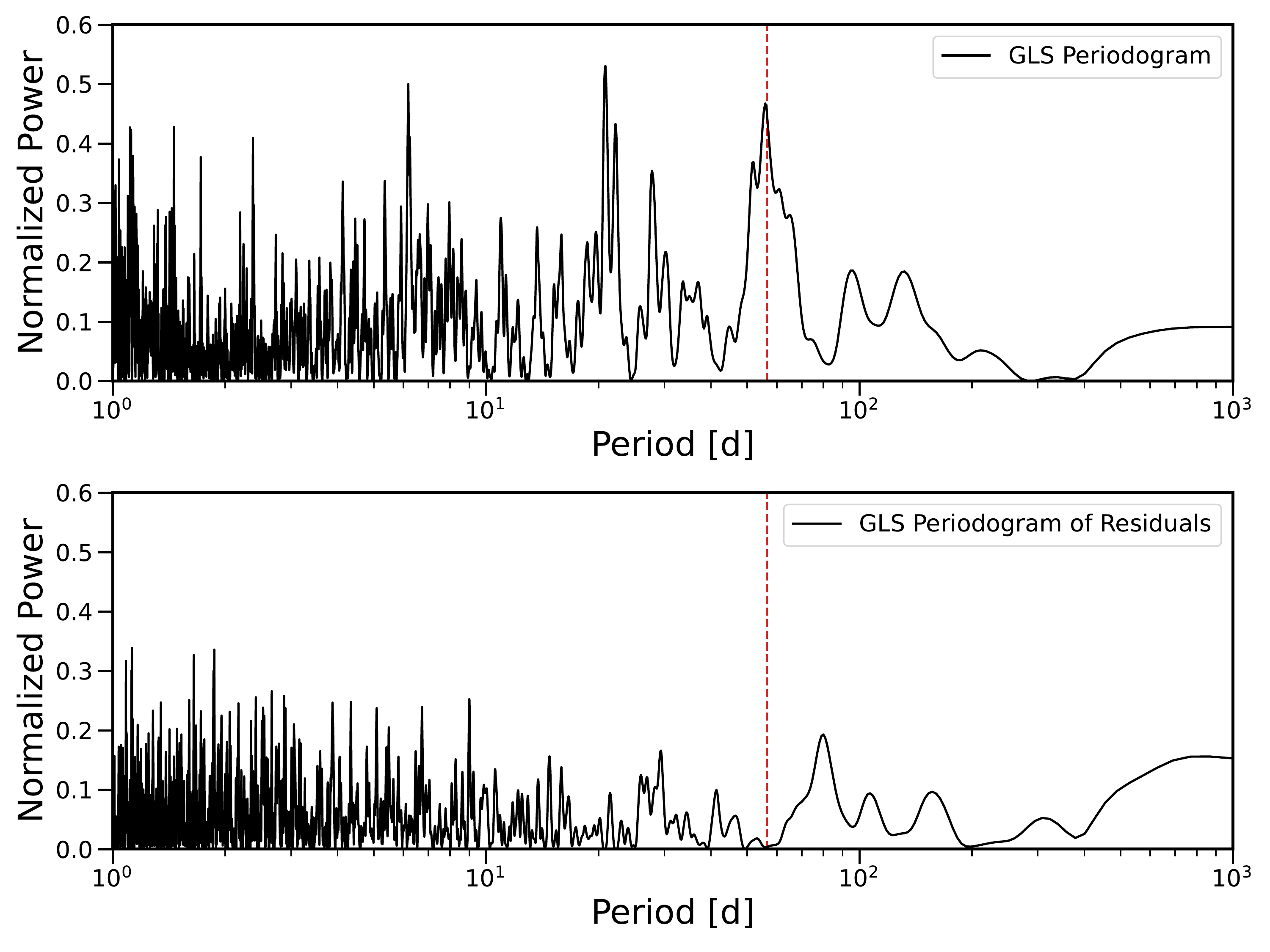}
    \caption{Generalized Lomb-Scargle periodograms for the full RV data set (top) and residuals (bottom). We mark the best-fit orbital period with a dashed red line. No periodic signals are prominent in the residuals.}
    \label{fig:periodogram}
\end{figure}

We compute the generalized Lomb-Scargle \citep{Zechmeister2009,Zechmeister2018gls} periodogram of the RV time series and residuals (residual RMS $= 13.8$ m s$^{-1}$) to explore whether the data might contain evidence for additional, non-transiting planets or other periodic signals (\autoref{fig:periodogram}). Following the removal of the $56.4$-day signal for TOI-4127 b, no significant periodic signals emerge. In addition, we find no evidence for any long term trends. The best-fit RV slope from the joint fit is $0.001\pm0.025$ m s$^{-1}$ d$^{-1}$, or $0.45\pm8.95$ m s$^{-1}$ yr$^{-1}$. We discuss additional constraints on the presence of external companions in Section \ref{sec:perturbers}.

\section{Discussion}\label{sec:discussion}

\subsection{Dynamical evolution of the TOI-4127 system}

\subsubsection{High-eccentricity tidal migration}

\begin{figure} 
    \centering
    \includegraphics[width=1.0\linewidth]{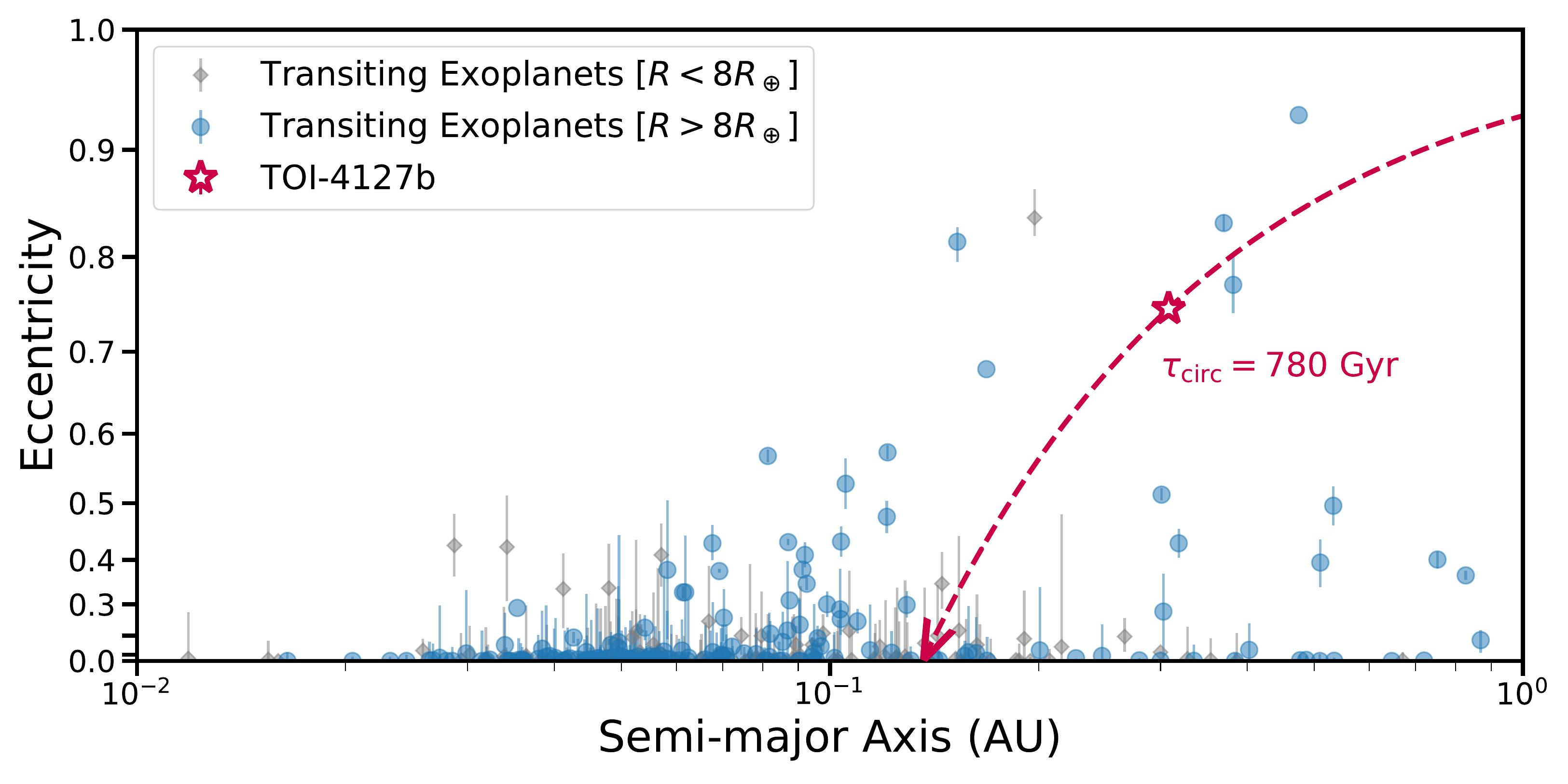}
    \caption{Eccentricities for all transiting exoplanets with $>3\sigma$ mass measurements. We show Jupiter-sized exoplanets ($R>8R_\oplus$) in blue and smaller planets in grey. TOI-4127 b (red) has an eccentricity higher than all but five exoplanets in this class. We also show the constant angular momentum tidal circularization track for this planet. TOI-4127 b is not expected to tidally circularize during the main sequence lifetime of its host star, although we note that the circularization timescale is highly sensitive to our choice of $Q_p$, which is not well constrained. Known planet parameters were taken from the NASA Exoplanet Archive on 2023 January 12.}
    \label{fig:eccentricity}
\end{figure}

The relatively high eccentricity of TOI-4127 b places it in a sparsely populated region of orbital parameter space\footnote{Based on data taken from the NASA Exoplanet Archive (\url{https://neid.ipac.caltech.edu/docs/NEID-DRP/}) on 2023 January 12}. Fewer than 10 other transiting warm Jupiters with $>3\sigma$ mass measurements have eccentricities $>0.5$ (\autoref{fig:eccentricity}). 
Of the four planets in this class with higher eccentricities than TOI-4127 b, two (HD 80606 b; \citeauthor{Naef2001} \citeyear{Naef2001} and TOI-3362 b; \citeauthor{Dong2021} \citeyear{Dong2021}) are expected to tidally circularize to become hot Jupiters, though this depends on other factors such as the rate of tidal dissipation in each system. Assuming a tidal dissipation efficiency parameter of $Q_p=10^{6.5}$ \citep{Jackson2008}, both of these exoplanets pass near enough to the tidal radii of their host stars at perisatron that tidal dissipation will be strong enough to induce inward migration.
To explore whether TOI-4127 b may also be a hot Jupiter progenitor en route to a tight, circular orbit, we first consider the timescale for tidal circularization. We calculate the circularization timescale as in Equations 2 and 3 of \citet{Adams2006},
\begin{equation}
\begin{split}
    \tau_{\rm circ} & \approx 1.6 {\rm\ Gyr}\left(\frac{Q_p}{10^6}\right)\left(\frac{m_p}{m_J}\right)\left(\frac{M_\star}{M_\odot}\right)^{-3/2}\\
    & \times \left(\frac{R_p}{R_J}\right)^{-5} \left(\frac{a}{0.05 {\rm\ AU}}(1-e^2)\right)^{13/2}[F(e^2)]^{-1}
\end{split}
\end{equation}
where we adopt the approximation $F(e^2)\approx1+6e^2+e^4$, dropping higher order terms. Again assuming $Q_p=10^{6.5}$, we find $\tau_{\rm circ}=780^{+280}_{-210}$ Gyr. Tidal dissipation would need to be two orders of magnitude more efficient than the value assumed here for the orbit to circularize before TOI-4127 evolves off the main sequence. This is plausible given that $Q_p$ is difficult to constrain for any individual system, as this parameter depends strongly on the physical parameters of the planet \citep[e.g.,][]{Ogilvie2004,Jackson2008}.

Apart from adopting a different value of $Q_p$, one way to significantly increase tidal dissipation efficiency is through tidal interactions that lead to energy exchange between the orbit and the exoplanet's \emph{f}-mode \citep{Wu2018}.
This effect requires that, at pericenter, the exoplanet comes within 4 times the tidal radius of the host star, $r_t$:
\begin{equation}
    a_{\rm peri} \leq 4 R_p(M_\star/M_p)^{1/3} .
\end{equation}
For TOI-4127 b, we find that $a_{\rm peri} \approx 18 r_t$, which is much too large for \emph{f}-mode dissipation to take effect.

We also consider the final orbital parameters of TOI-4127 b in the case that the orbit does ultimately circularize. If we assume that the angular momentum of the exoplanet is conserved during high-eccentricity tidal migration (i.e., no mass loss and no angular momentum exchange), the final semi-major axis is given by
\begin{equation}
   a_{\rm present}(1-e_{\rm present}^2) = a_{\rm final}(1-e_{\rm final}^2) ,
\end{equation}
and for a circular orbit with $e_{\rm final}=0$, $a_{\rm final}=0.1361^{+0.0050}_{-0.0046}$ AU. Both the final semi-major axis and the corresponding orbital period ($P_{\rm final}=16.56^{+0.72}_{-0.65}$ days) are inconsistent with the population of hot Jupiters, which exhibit a pile-up at semi-major axes $a<0.1$ AU and orbital periods of 3--4 days \citep{Wright2009,Santerne2016}.

TOI-4127 b does not presently have the characteristics of a hot Jupiter progenitor in the same class as HD 80606 b and TOI-3362 b. This will remain true if the present eccentric orbit is the result of a one-time excitation event with no avenue for future angular momentum exchange, such as planet-planet scattering \citep{Rasio1996} in which the second body is ejected from the system. As \citet{Ford2001} show, a scattering event in which one planet is ejected can excite the eccentricity of the retained planet up to $e_{\rm max}\approx 0.8$ if the planets are of equal mass. This scenario could plausibly explain the observed eccentricity of TOI-4127 b ($e=0.7471^{+0.0078}_{-0.0086}$).
However, the high eccentricity may instead be a product of ongoing interactions with an external perturber, such as perturber-coupled eccentricity oscillation or a series of planet-planet scattering events. If a distant, massive perturber is present in a system with a warm Jupiter, angular momentum exchange with this perturber may periodically place the inner planet on a high-eccentricity tidal migration track \citep{Dong2014}.

\subsubsection{Constraints on the presence of an undetected perturber}\label{sec:perturbers}

We do not detect any additional stellar or planetary companions to TOI-4127, but coupling with an undetected perturber might still be responsible for the high eccentricity of the detected warm Jupiter. We show in Section \ref{sec:nessi} using NESSI speckle imaging data that TOI-4127 lacks any bright neighbors, but the NESSI data alone do not constrain the presence of low mass companions such as mid- to late-K dwarfs and M dwarfs and substellar objects. We consider the detectability of such an object given the available data here. First, we note that Gaia does not detect any sources within 25'' (projected separation $\approx6800$ AU) of TOI-4127 \citep{GaiaCollaboration2022}, which is consistent with the absence of sources detected with NESSI and the lack of contamination observed in the pixel-level TESS light curves. In addition, the Gaia astrometric solution for this star has a renormalized unit weight error (RUWE) of 1.27 \citep{GaiaCollaboration2022}. The RUWE quantifies how well the solution conforms to that of a single star; while values significantly larger than 1 indicate a preference for a multiple object astrometric solution, \citet{Penoyre2020} suggest a threshold of RUWE$>1.4$ for this to be attributed to orbital motion induced by a bound companion.

Following the methods of \citet{Jackson2021}, which we summarize below, we next construct a population of perturbers capable of exciting eccentricity oscillations in TOI-4127 b and facilitating migration to a final semi-major axis of $a_{\rm final}<0.1$ AU, and we then assess the detectability of members of this population given the available observational data.
We only consider radial velocity and astrometric detectability \rrev{in full, and we comment briefly on transit timing variations.} With only three transits, the existing photometry is insufficient \rstrike{to meaningfully evaluate any} \rrev{for robust analysis of} transit timing variations or transit duration variations for this system.

We first build a population of companion planets capable of inducing strong eccentricity oscillations in TOI-4127 b by drawing planetary masses and orbital parameters from a set of analytical distributions informed by observations of long-period giant planets. 
Although the population of long-period giants is not well-constrained by observations, \citet{Jackson2021} show that their results are robust to different choices for the underlying companion distributions.
For this work, we draw masses from a power-law distribution set by \citet{Cumming2008} over the range of 0.1 to 20 $M_J$, orbital periods from a broken power-law distribution set by \citet{Fernandes2019} over the range of 200 to 100,000 days with a turnover in occurrence rate at 859 days, eccentricities from a beta distribution ($\alpha = 0.74$, $\beta = 1.61$) fit to the set of confirmed long-period giant planets on the NASA Exoplanet Archive, inclinations from an isotropic distribution, and the remaining orbital angles from uniform distributions. 
In order for perturber-coupled high-eccentricity migration to occur in this system, TOI-4127 b and the undetected companion must be long-term stable and the precession induced by the outer companion must be much faster than general relativistic precession. We apply these two analytical cuts to our simulated population of potential perturbers following Section 2.4 of \citet{Jackson2021}. 

Next, we assess the detectability of each perturber by modeling the two-planet RV signal of the system. 
We draw planetary parameters for TOI-4127 b from the best-fit posterior distribution of the joint transit + RV fit in Section \ref{sec:jointfit}, and we 
randomly draw a companion from our simulated perturber population.
Using the NEID and SOPHIE observation times and measurement uncertainties, we calculate the RV signal that would have been observed for this two-planet system.
We then fit and subtract the TOI-4127 b signal using the \textsc{mpfit} \textsc{idl} package \citep{Markwardt2009}, fixing the orbital period and time of conjunction, leaving us with the residual signal from the perturbing companion. We repeat this process for 1000 perturbers and then determine the detectability of each residual signal by measuring the ratio of the slope of the residual RV signal to the estimated error in the slope ($|\rm{slope}|/e_{\rm{slope}}$). As in \citet{Jackson2021}, we consider the companion to be detectable for each trial in which this metric is greater than 3.5 over the full RV baseline. To assess sensitivity to perturbers with orbital periods shorter than the full RV baseline, we also calculate the slope ratio for each of the following limited baselines: (1) the full NEID baseline, including all overlapping SOPHIE measurements, (2) the full SOPHIE baseline, including all overlappind NEID measurements, and (3) BJD 2459450 through BJD 2459700, i.e., all data excluding the earliest NEID measurements and latest SOPHIE measurements.

Perturbing companions may also be detectable with Gaia astrometry. For each of the above trials, we model the transverse motion of the stellar host due to the gravitational pull of the perturber following \citet{Quirrenbach2010} and subtract off a linear fit to account for proper motion.
For each trial, we sample the model at 473 points, corresponding to the number of measurements used in the Gaia Data Release 3 astrometric solution \citep{GaiaCollaboration2022} for TOI-4127, and we calculate the maximum angular distance, $\Delta\theta$, between model samples. We compare this value to a conservative detection limit of $100\mu\rm{as}$ \citep{Perryman2014}, and we label trials with $\Delta\theta$ above this limit as detectable via Gaia astrometry.

\begin{figure} 
    \centering
    \includegraphics[width=1.0\linewidth]{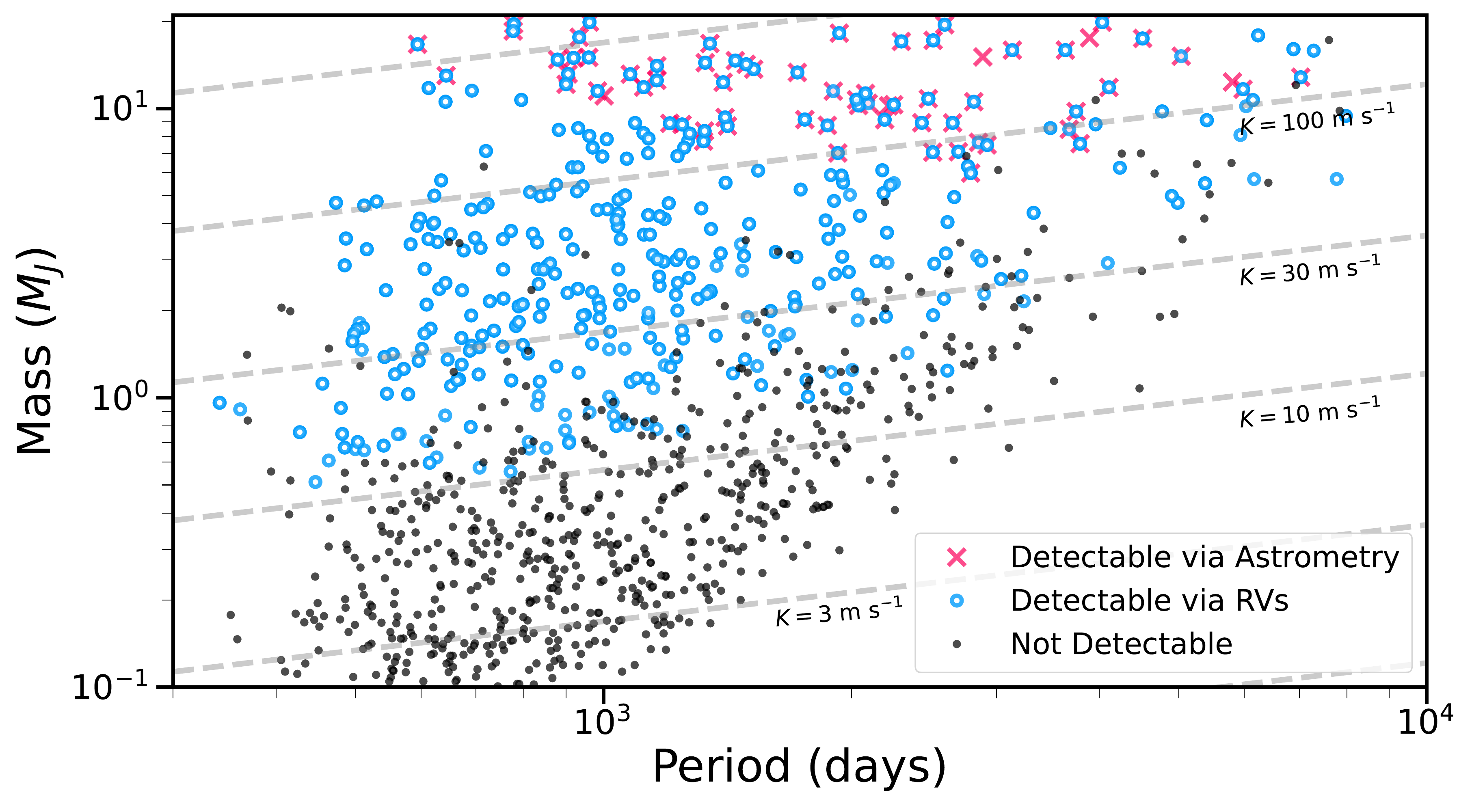}
    \caption{Simulated population of long-period perturbers capable of exciting the eccentricity of TOI-4127 b to facilitate high-eccentricity tidal migration. We show the perturbers that are detectable via available astrometric and radial velocity observations as pink `x' marks and blue circles, respectively, and the perturbers that cannot yet be detected as black points.}
    \label{fig:perturbers}
\end{figure}

We show the sample of potential perturbers in \autoref{fig:perturbers}. While signals from most of these objects with RV semi-amplitudes greater than $\sim10$ m s$^{-1}$ would have been detected in the available RV or astrometric data sets, exoplanets with masses $K<10$ m s$^{-1}$ are overwhelmingly not detectable.
We cannot rule out perturber-coupled eccentricity oscillations as the source of the eccentric orbit of TOI-4127 b and as a mechanism for future high-eccentricity tidal migration. In addition, given that a fraction of potential perturbers shown in \autoref{fig:periodogram} have $K<3$ m s$^{-1}$, it is unlikely that RV observations alone could definitely confirm or reject the perturber-coupled eccentricity oscillation scenario without increased significant observational expense. At the reported NEID single measurement precision of $6.13$ m s$^{-1}$ for this target, one would need more than 30 additional observations to achieve even a $3\sigma$ mass measurement of a perturber with $K=3$ m s$^{-1}$.

\rrev{To determine whether transit timing variations might be detectable, we fit the observed transit midpoints and we compare these to the expected transit ephemeris for each of the three transits based on the best-fit period and time of conjunction. We find that the transit midpoints deviate from the best fit ephemeris by $135\pm107$ seconds, $-246\pm136$ seconds, and $125\pm110$ seconds for the transits in Sectors 20, 26, and 53, respectively. Given the corresponding ephemeris uncertainties of $\pm86$, $\pm69$, and $\pm104$ seconds, we find no evidence for significant timing variations.}

\subsection{Prospects for atmospheric characterization}

Under the na\"ive assumption that the planetary surface temperature responds instantaneously to changing irradiation during its elliptical orbit, the equilibrium surface temperature of TOI-4127 b during transit is $T_{\rm eq, tr}=1145^{+30}_{-29}$ K for a surface albedo of 1, and the corresponding transmission spectroscopy metric \citep[TSM;][]{Kempton2018} is  $15.3^{+1.5}_{-1.3}$. On its own, this modest TSM does not make TOI-4127 b an exceptional target for atmospheric characterization with JWST, as it falls well below the threshold of TSM$>96$ recommended by \citet{Kempton2018}.
However, phase curve observations across different parts of the orbit are considerably more compelling. The high orbital eccentricity will lead to extreme temperature variations, exceeding $1200$ K at pericenter and dropping below $500$ K at apocenter, and potentially changes in the equilibrium chemistry of the atmosphere \citep{Fortney2020,Mayorga2021}. Even a pair of measurements during transit and occultation ($T_{\rm eq, oc} = 590^{+15}_{-13}$ K), if feasible, would nearly capture the two temperature extremes and serve as a powerful probe of the atmospheric chemistry of TOI-4127 b.
We use the fitted orbital parameters to calculate the secondary impact parameter as 
\begin{equation}
    b_2 = b\frac{1+e\sin(\omega)}{1-e\sin(\omega)}=0.65^{+0.51}_{-0.43}
\end{equation}
and find that an occultation is very likely, though not guaranteed.
Phase curve measurements can also be used to place constraints on the heat redistribution timescale and thus rotation period of TOI-4127 b \citep[e.g.,][]{Lewis2014,deWit2016,Lewis2017}. In contrast to the fully tidally-locked synchronous rotation of short-period planets with circular orbits, eccentric warm Jupiters are expected to exhibit pseudo-synchronous rotation, in which their rotation rates are approximately equivalent to the orbital frequency at periastron \citep{Hut1981,Ivanov2007}. Deviation from (or adherence to) pseudo-synchronous rotation can shed light on the tidal interaction timescale of the system.

\subsection{Rossiter-McLaughlin effect}

\begin{figure} 
    \centering
    \includegraphics[width=1.0\linewidth]{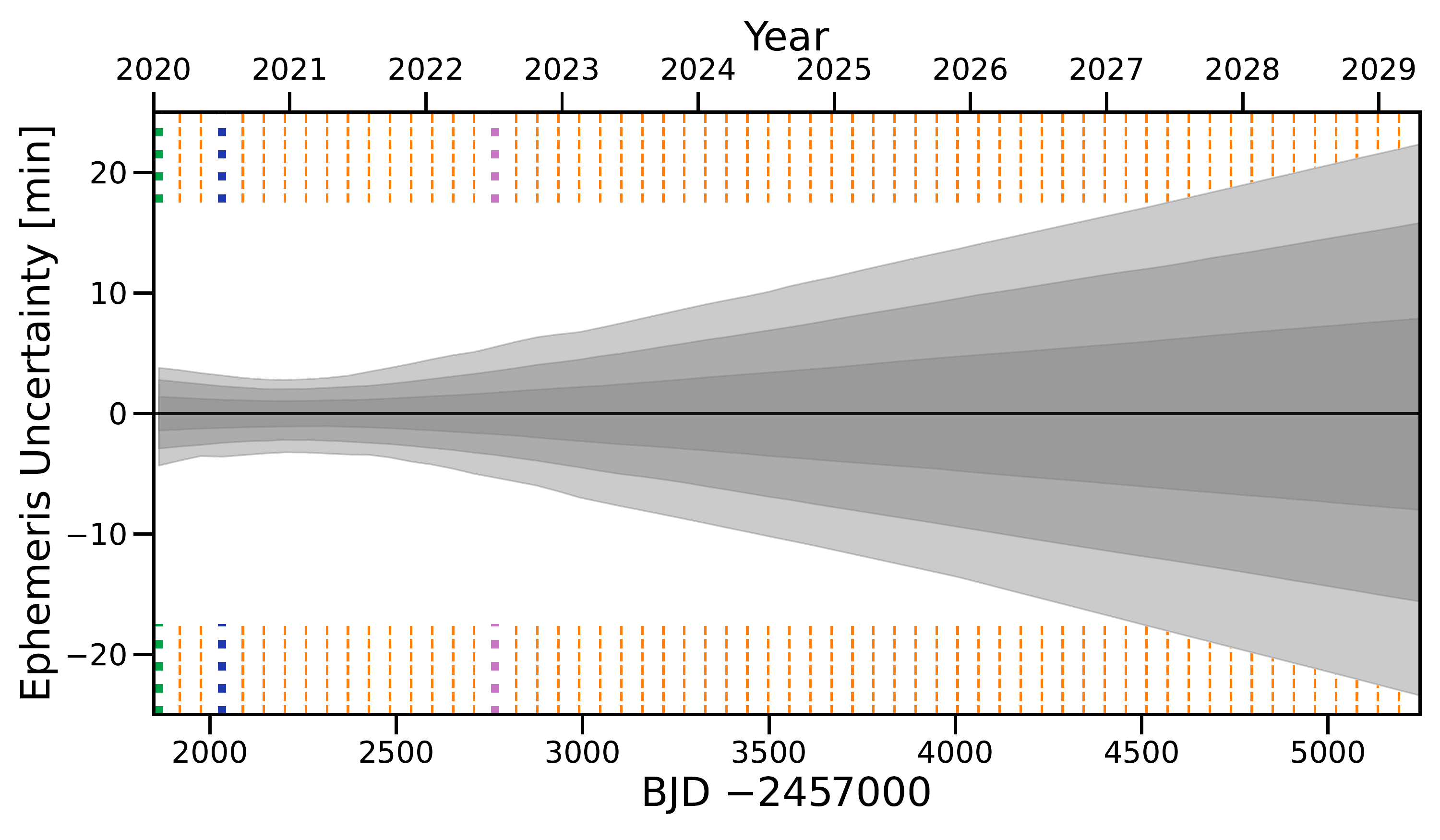}
    \caption{Projected transit ephemeris for TOI-4127 b for the 60 orbits following the transit detected in Sector 20. The $1$, $2$, and $3$ sigma uncertainties on the ephemeris are shown as progressively lighter, grey shaded regions. Past and future transit times are indicated by vertical orange bars, and the observed transits in Sectors 20, 26, and 53 are indicated in green, blue, and pink, respectively, to match the light curves in \autoref{fig:tess_lc}. We expect the ephemeris to remain precise to better than 25 minutes, or 10\% of the transit duration, through early 2029.}
    \label{fig:ephemeris}
\end{figure}

Using the fitted orbital period and time of conjunction, we estimate the projected uncertainty on the TOI-4127 b transit ephemeris for the next 40 orbits ($\sim6$ years). As shown in \autoref{fig:ephemeris}, the ephemeris is expected to remain precise to better than 25 minutes, or $\sim10\%$ of the transit duration, over this time frame. This is sufficient not only for the planning and execution of atmospheric measurements during transit and eclipse, but also for measurements of the stellar obliquity via the Rossiter-McLaughlin (RM) effect \citep{Rossiter1924,McLaughlin1924}.

Recent work by \citet{Rice2022} has shown that warm Jupiters are preferentially more aligned than hot Jupiters. These findings have implications for the formation histories and dynamical evolution of these systems \citep{Wang2021,Rice2022hj}.
But the \citet{Rice2022} sample includes only two warm Jupiters orbiting host stars with effective temperatures above the Kraft break \citep{Kraft1967}, defined as $T_{\rm eff}\simeq 6100$ K, which is important given observed correlations between alignment and stellar temperature \citep{Albrecht2012}.
Measurement of the RM effect for TOI-4127, which has an effective temperature of $T_{\rm eff}=6096\pm115$ K, would add a valuable data point to population-level studies of stellar obliquities. 

We calculate the expected semi-amplitude of the RM signal as
\begin{equation}
    A_{\rm RM} = \frac{2\delta}{3} v\sin i \sqrt{1-b^2}
\end{equation}
which yields $37.7\pm1.3$ m s$^{-1}$ for \rstrike{TOI-4127} \rrev{the $v\sin i$ value of $7.7$ km s$^{-1}$ from SPC analysis of TRES data}.\rstrike{The projected rotational velocity, $v\sin i$, was}\rstrike{ estimated to be $7.7$ km s$^{-1}$ based on analysis of data  }\rstrike{ from the Tillinghast Reflector Echelle Spectrograph }\rstrike{ (TRES) using the Stellar Parameter Classification Tool} \rstrike{ \citep[SPC;][]{Buchhave2012,Buchhave2014}.} \rrev{Measurement of} a RM signal of this magnitude would be feasible with an instrument such as NEID, for which we achieve $\sim6$ m s$^{-1}$ RV precision on this target with 15 minute exposures. However, given the long orbital period of TOI-4127 b, opportunities to observe a complete transit from the ground will be relatively rare. Using the projected transit ephemerides shown in \autoref{fig:ephemeris}, we identify the set of viable RM opportunities through the end of 2028 for several ground-based telescopes and instruments with the requisite aperture and precision RV measurement capabilities. In addition to NEID and SOPHIE, we consider the Keck Planet Finder \citep[KPF;][]{Gibson2016} and MAROON-X \citep{Seifahrt2018}, both on Maunakea, and the HARPS-N spectrograph \citep{Cosentino2012} on the Telescopio Nazionale Galileo on La Palma Island. Here, we define a viable opportunity to be one for which the solar altitude is $<-18^\circ$ and the target airmass is $<2$ for the duration of the transit; the list of transit dates is shown in \autoref{tab:rm}.

\begin{deluxetable}{lll}
\tablecaption{Ground-based Rossiter-McLaughlin Measurement Opportunities for TOI-4127 b \label{tab:rm}}
\tablehead{\colhead{UTC Date}&  \colhead{Transit Midpoint (BJD)}&
\colhead{Instrument}}
\startdata
2023 November 24 & $2460272.6974^{+0.0020}_{-0.0021}$ & HARPS-N \\
2024 March 15 & $2460385.4950^{+0.0022}_{-0.0022}$ & HARPS-N, \\
\textemdash & \textemdash & SOPHIE \\
2024 December 22 & $2460667.4888^{+0.0026}_{-0.0027}$ & HARPS-N \\
2025 September 30 & $2460949.4827^{+0.0032}_{-0.0032}$ & SOPHIE \\
2025 November 26 & $2461005.8814^{+0.0033}_{-0.0033}$ & NEID, \\
\textemdash & \textemdash & SOPHIE \\
2026 December 26 & $2461400.6729^{+0.0039}_{-0.0040}$ & NEID, \\
\textemdash & \textemdash & HARPS-N, \\
\textemdash & \textemdash & SOPHIE \\
2027 April 17 & $2461513.4704^{+0.0041}_{-0.0042}$ & SOPHIE \\
2027 October 4 & $2461682.6667^{+0.0045}_{-0.0045}$ & HARPS-N \\
2027 November 29 & $2461739.0655^{+0.0046}_{-0.0047}$ & KPF, \\
\textemdash & \textemdash & MAROON-X \\
2028 January 24 & $2461795.4642^{+0.0047}_{-0.0048}$ & HARPS-N, \\
\textemdash & \textemdash & SOPHIE \\
2028 November 1 & $2462077.4581^{+0.0052}_{-0.0053}$ & SOPHIE \\
2028 December 28 & $2462133.8569^{+0.0053}_{-0.0054}$ & NEID \\
\enddata
\end{deluxetable}

\section{Summary}
In this work, we present the discovery of the highly eccentric warm Jupiter TOI-4127 b with TESS, and the confirmation and characterization of the exoplanet signal with the NEID and SOPHIE spectrographs. We jointly fit the transit and RV data to constrain the physical ($R_p = 1.096^{+0.039}_{-0.032} R_J$,  $M_p = 2.30^{+0.11}_{-0.11} M_J$) and orbital ($P = 56.39879^{+0.00010}_{-0.00010}$ d, $e=0.7471^{+0.0078}_{-0.00086}$) parameters of the exoplanet. The high orbital eccentricity makes this rare system a compelling target for dynamical studies.
We comment on the potential for dynamical evolution of TOI-4127 b in the context of hot Jupiter formation pathways, and we find that TOI-4127 b may indeed be a hot Jupiter progenitor, but only if a perturbing companion is present.
While we find no evidence for additional bodies in this system, we show that an undetected perturber could be capable of exciting eccentricity oscillations and high-eccentricity migration. Future RV follow-up observations with higher measurement precision could shed light on this possibility. We also comment on the feasibility of measuring the stellar obliquity of TOI-4127 with Rossiter-McLaughlin observations, and on the importance of such a measurement for population-level studies of warm Jupiter host stars.

\section{Acknowledgements}

NEID is funded by NASA through JPL by contract 1547612 and the NEID Data Reduction Pipeline is funded through JPL contract 1644767. Funding for this work was partially provided by Research Support Agreements 1646897 and 1679618 administered by JPL.
The Center for Exoplanets and Habitable Worlds and the Penn State Extraterrestrial Intelligence Center are supported by the Pennsylvania State University and the Eberly College of Science. GS acknowledges support provided by NASA through the NASA Hubble Fellowship grant HST-HF2-51519.001-A awarded by the Space Telescope Science Institute, which is operated by the Association of Universities for Research in Astronomy, Inc., for NASA, under contract NAS5-26555. PD acknowledges support by a 51 Pegasi b Postdoctoral Fellowship from the Heising-Simons Foundation. PCZ acknowledges funding from the French National Research Agency (ANR) under contract number ANR-18-CE31-0019 (SPlaSH).
DD acknowledges support from the NASA Exoplanet Research Program grant 18-2XRP18\_2-0136.
This research has made use of the SIMBAD database, operated at CDS, Strasbourg, France, and NASA's Astrophysics Data System Bibliographic Services.
Part of this work was performed for the Jet Propulsion Laboratory, California Institute of Technology, sponsored by the United States Government under the Prime Contract 80NM0018D0004 between Caltech and NASA. This work is partly supported by the French National Research Agency in the framework of the Investissements d’Avenir program (ANR-15-IDEX-02), through the funding of the "Origin of Life" project of the Grenoble-Alpes University This work has been partly supported by a grant from Labex OSUG at 2020 (Investissements d’avenir – ANR10 LABX56). This work was supported by the ``Programme National de Plan\'etologie'' (PNP) of CNRS/INSU and the French Space Agency (CNES).
The results reported herein benefitted from collaborations and/or information exchange within NASA’s Nexus for Exoplanet System Science (NExSS) research coordination network sponsored by NASA’s Science Mission Directorate under Agreement No. 80NSSC21K0593 for the program ``Alien Earths''.

This paper contains data taken with the NEID instrument, which was funded by the NASA-NSF Exoplanet Observational Research (NN-EXPLORE) partnership and built by Pennsylvania State University. NEID is installed on the WIYN telescope, which is operated by the NSF's National Optical-Infrared Astronomy Research Laboratory, and the
NEID archive is operated by the NASA Exoplanet Science Institute at the California Institute of Technology.
Some of the observations in this paper made use of the NN-EXPLORE Exoplanet and Stellar Speckle Imager (NESSI). NESSI was funded by the NASA Exoplanet Exploration Program and the NASA Ames Research Center. NESSI was built at the Ames Research Center by Steve B.\ Howell, Nic Scott, Elliott P.\ Horch, and Emmett Quigley.
NN-EXPLORE is managed by the Jet Propulsion Laboratory, California Institute of Technology under contract with the National Aeronautics and Space Administration. We thank the NEID Queue Observers and WIYN Observing Associates for their skillful execution of our observations. We also thank the Observatoire de Haute-Provence (CNRS) staff for their support in obtaining SOPHIE data. 

This work includes data collected by the TESS mission, which are publicly available from MAST. Funding for the TESS mission is provided by the NASA Science Mission directorate. We acknowledge the use of public TESS data from pipelines at the TESS Science Office and at the TESS Science Processing Operations Center. Resources supporting this work were provided by the NASA High-End Computing (HEC) Program through the NASA Advanced Supercomputing (NAS) Division at Ames Research Center for the production of the SPOC data products. This research has made use of the Exoplanet Follow-up Observation Program website, which is operated by the California Institute of Technology, under contract with the National Aeronautics and Space Administration under the Exoplanet Exploration Program. Some of the data presented in this paper were obtained from MAST. Support for MAST for non-HST data is provided by the NASA Office of Space Science via grant NNX09AF08G and by other grants and contracts. This work has made use of data from the European Space Agency (ESA) mission {\it Gaia} (\url{https://www.cosmos.esa.int/gaia}), processed by the {\it Gaia} Data Processing and Analysis Consortium (DPAC, \url{https://www.cosmos.esa.int/web/gaia/dpac/consortium}). Funding for the DPAC has been provided by national institutions, in particular the institutions participating in the {\it Gaia} Multilateral Agreement. This research has made use of the NASA Exoplanet Archive, which is operated by the California Institute of Technology, under contract with the National Aeronautics and Space Administration under the Exoplanet Exploration Program.

This research made use of \textsf{exoplanet} \citep{Foreman-Mackey2021,exoplanet:zenodo} and its dependencies \citep{Foreman-Mackey2017,Foreman-Mackey2018,Agol2020,Kumar2019,exoplanet:astropy13,AstropyCollaboration2018,Kipping2013,Salvatier2016,exoplanet:luger18}.
Based in part on observations at Kitt Peak National Observatory, NSF’s NOIRLab (Prop. ID 2021A-0388 \& 2021B-0432; PI: A.\ Gupta), managed by the Association of Universities for Research in Astronomy (AURA) under a cooperative agreement with the National Science Foundation. The authors are honored to be permitted to conduct astronomical research on Iolkam Du’ag (Kitt Peak), a mountain with particular significance to the Tohono O’odham.
We also express our deepest gratitude to Zade Arnold, Joe Davis, Michelle Edwards, John Ehret, Tina Juan, Brian Pisarek, Aaron Rowe, Fred Wortman, the Eastern Area Incident Management Team, and all of the firefighters and air support crew who fought the recent Contreras fire. Against great odds, you saved Kitt Peak National Observatory.

The Pennsylvania State University campuses are located on the original homelands of the Erie, Haudenosaunee (Seneca, Cayuga, Onondaga, Oneida, Mohawk, and Tuscarora), Lenape (Delaware Nation, Delaware Tribe, Stockbridge-Munsee), Shawnee (Absentee, Eastern, and Oklahoma), Susquehannock, and Wahzhazhe (Osage) Nations.  As a land grant institution, we acknowledge and honor the traditional caretakers of these lands and strive to understand and model their responsible stewardship. We also acknowledge the longer history of these lands and our place in that history.

\facilities{TESS, WIYN (NEID, NESSI), SOPHIE, \textit{Gaia}}

\software{\textsf{astropy} \citep{AstropyCollaboration2018}, \textsf{arviz} \citep{Kumar2019}, \textsf{barycorrpy} \citep{Kanodia2018}, \textsf{eleanor} \citep{Feinstein2019}, \textsf{lightkurve} \citep{LightkurveCollaboration2018}, \textsf{matplotlib} \citep{Hunter2007}, \textsf{MPFIT} \citep{Markwardt2009}, \textsf{numpy} \citep{Harris2020}, \textsf{PyMC3} \citep{Salvatier2016}, \textsf{scipy} \citep{Oliphant2007}, \textsf{SERVAL} \citep{Zechmeister2018} \textsf{Theano} \citep{TheTheanoDevelopmentTeam2016}}

\bibliography{references}{}
\bibliographystyle{aasjournal}

\end{document}